\documentclass[preprintnumbers,article,amsmath,amssymb,floatfix,10pt,prd,superscriptaddress,nofootinbib,showkeys]{revtex4}

\usepackage{doi}%<----------
\usepackage{hyperref}
\hypersetup{
	colorlinks=true,        % false: boxed links; true: colored links
	linkcolor=red,         % color of internal links
	citecolor=cyan,         % color of links to bibliography
}

\usepackage{graphicx}% Include figure files
\usepackage{dcolumn}% Align table columns on decimal point
\usepackage{bm}% bold math
\usepackage{color}
\usepackage{enumitem}
\usepackage{amsmath}
\usepackage{amssymb}

\newcommand{\beq}{\begin{equation}}
\newcommand{\eeq}{\end{equation}}
\newcommand{\bea}{\begin{eqnarray}}
\newcommand{\eea}{\end{eqnarray}}

\usepackage{bbm}
\usepackage{amsfonts}
\usepackage{mathrsfs}
\usepackage{latexsym}
\usepackage{epsfig}
\usepackage{epstopdf}
\usepackage{epstopdf}
\usepackage{graphicx}
\usepackage{amssymb}
\usepackage{amsmath}
\usepackage{dcolumn}
\usepackage{bm}
\usepackage{color}
\usepackage{comment}
\usepackage{float}
\usepackage{xcolor}
\usepackage{orcidlink}

\begin{document}

\preprint{}

\title{Strong Deflection Gravitational Lensing by Non-Exotic Traversable Wormholes in King and Dekel-Zhao Dark Matter Halos under $f(R, L_m)$ Gravity}

\author{Susmita Sarkar\,\orcidlink{0009-0007-1179-2495}}
\email{susmita.mathju@gmail.com}
\affiliation{Department of Applied Science and Humanities, Haldia Institute of Technology, Haldia 721606, West Bengal, India}

\author{Nayan Sarkar\,\orcidlink{0000-0002-3489-6509}}
\email{nayan.mathju@gmail.com}
\affiliation{Department of Mathematics, Karimpur Pannadevi College, Nadia 741152, West Bengal, India}

\author{Abdelmalek Bouzenada\,\orcidlink{0000-0002-3363-980X}}
\email{abdelmalekbouzenada@gmail.com (Corresp. author)}
\affiliation{Laboratory of Theoretical and Applied Physics, Echahid Cheikh Larbi Tebessi University, 12001 Tebessi, Algeria}
\affiliation{Research Center of Astrophysics and Cosmology, Khazar University, Baku AZ1096, 41 Mehseti Street, Azerbaijan}

\author{Anikul Islam\,\orcidlink{0009-0001-4929-5298}}
\email{anikulislam0025@gmail.com}
\affiliation{Department of Mathematics, Jadavpur University, Kolkata 700032, West Bengal, India}

\author{Farook Rahaman\,\orcidlink{0000-0003-0594-4783}}
\email{rahaman@iucaa.ernet.in}
\affiliation{Department of Mathematics, Jadavpur University, Kolkata 700032, West Bengal, India}

\date{\today}

\begin{abstract}
In this study, we analyze asymptotically flat non-exotic traversable wormhole geometries embedded in the King and Dekel-Zhao dark matter halos within the framework of $f(R, L_{m})$ gravity. Two functional realizations of the theory are adopted: \textbf{Model-I:} $f(R, L_{m})=(R/2)+L_{m}^{\alpha}$ and \textbf{Model-II:} $f(R, L_{m})=(R/2)+(1+\lambda R)L_{m}$. For both models, wormhole solutions are constructed and examined using the King and Dekel–Zhao dark matter density profiles, allowing for an assessment of how the underlying matter distribution influences the wormhole structure. The energy conditions are evaluated to confirm the possibility of supporting the wormhole geometries with non-exotic matter, while embedding surfaces, proper radial distance, and total gravitational energy are investigated to characterize physical viability and traversability. Furthermore, we analyze the strong deflection angle and its consequences for gravitational lensing and identify potential observational signatures associated with these wormhole configurations. In this case, the results demonstrate that, in $f(R, L_m)$ gravity, for suitable parameter selections, dark matter environments can sustain physically consistent, non-exotic traversable wormhole geometries with characteristic gravitational lensing signatures, offering insight into the coupling between modified gravity, dark matter, and astrophysical observations.
\end{abstract}


\maketitle

%%%%%%%%%%%%%%%%%%%%%%%%%%%%%%%%%%%%%%%%%%%%%%
\section{Introduction}\label{sec1}
%%%%%%%%%%%%%%%%%%%%%%%%%%%%%%%%%%%%%%%%%%%%%%

In contemporary theoretical astrophysics, wormholes are regarded as highly intriguing structures that naturally emerge from the framework of Einstein’s general relativity (GR) when special kinds of matter distributions are considered. In this case, these hypothetical objects can serve as tunnels or conduits that connect either widely separated regions of the same universe or even two different universes. The first notion of such a tunnel-like configuration was introduced by Flamm \cite{POT1}, who analyzed the Schwarzschild solution in 1916. Later, Einstein and Rosen \cite{POT2} advanced this perspective by presenting a bridge-type connection between two distinct external spacetime regions, leading to the concept of the Einstein–Rosen bridge (ERB). Also, wheeler subsequently coined the term wormhole \cite{POT3}, and together with Fuller, they illustrate that these structures are dynamically unstable and prone to collapse after their creation, which rules out their utility for interstellar travel in that form \cite{POT4}. Furthermore, the interest in traversable Lorentzian wormholes reignited when Morris and Thorne \cite{POT5} showed that a spherically symmetric wormhole geometry can, in principle, provide a two-way passage between two asymptotically flat spacetimes, provided that exotic matter is present at the throat to keep the passage open. Such constructions necessarily lead to violations of the null energy condition (NEC), raising the question of whether exotic matter might arise in nature within the framework of quantum field theory \cite{POT7}. Visser’s monograph \cite{POT6} and later work with collaborators \cite{POT8} illustrate that specific spacetime geometries could sustain traversable wormholes with a minimized requirement of exotic matter, thereby enhancing their physical plausibility. Alongside these efforts, the development of modified theories of gravity has provided powerful new approaches for studying wormholes, motivated by the fact that GR with ordinary matter alone cannot explain phenomena such as the late-time acceleration of the universe or the presence of dark matter (DM). Another important result in this modified framework, including $f(R)$ gravity \cite{POT9,POT10,POT11}, $f(R,T) $ gravity \cite{POT12,POT13,POT14}, brane-world models \cite{POT15,POT16,POT17}, Rastall gravity \cite{POT18,POT19,POT20}, and $f(Q)$ gravity \cite{POT21,POT22,POT23,POT24}, have been employed to construct wormhole solutions by assuming or deriving suitable shape function models, with detailed examinations of stability and energy conditions. More recently, non-linear $f(R,L_m)$ models have been studied in the context of traversable wormholes \cite{POT25}, and further works have investigated their compatibility with various energy conditions \cite{POT26}. In this context, additional important contributions exploring wormhole geometries within modified gravity frameworks are available in the literature \cite{POT27,POT28,POT29,POT30}.

Modified gravity theories were illustrated by Harko et al. \cite{FLMG1}, who introduced the $f(R, L_m)$ framework as a generalization of the $f(R)$ models. In this formulation, $R$ defines the Ricci scalar curvature that encodes the geometric features of spacetime, while $L_m$ symbolizes the matter Lagrangian density, thereby combining the geometrical attributes of gravity with the material distribution of matter. Also, the relation between these two elements introduces a non-trivial coupling, giving rise to an additional force orthogonal to the particle’s four-velocity, which in turn drives massive particles away from purely geodesic trajectories. Also, this property distinguishes $f(R, L_m)$ gravity from conventional relativistic dynamics and opens avenues for reinterpreting the motion of particles and fields in curved spacetime. Following its proposal, subsequent investigations have broadened the scope of this idea by analyzing arbitrary couplings between matter and geometry \cite{FLMG2}, thereby deepening our understanding of both astrophysical and cosmological processes. In this context, many studies have tested its phenomenological implications, particularly in relation to the evolution of the universe, cosmic structure formation, and modifications to astrophysical systems \cite{FLMG3, FLMG4, FLMG5, FLMG6, FLMG7}. However, a central feature of this theory is its violation of the equivalence principle, which is subject to strong experimental constraints derived from solar system observations \cite{FLMG8, FLMG9}. In the cosmological context, Wang and Liao examined the validity of energy conditions within $f(R, L_m)$ gravity, thereby providing important insights into the theoretical consistency of the model \cite{FLMG10}. In a complementary direction, Gonçalves and Moraes are testing cosmological dynamics with non-minimal matter-geometry couplings \cite{FLMG11}, while Solanki et al. show how the presence of bulk viscosity in an anisotropic background could lead to late-time cosmic acceleration \cite{FLMG12}. In this case, for more discussion and expalin, Jaybhaye et al. tested viscous dark energy models within this framework and derived constraints on the equation of state parameter, illustrating new perspectives on the driving forces behind cosmic expansion \cite{FLMG13}. Also, these efforts show and explain the phenomenology of $f(R, L_m)$ gravity models, where the synergy between matter and geometry reshapes fundamental principles and yields novel interpretations of dark energy, acceleration, and the large-scale dynamics of the universe. In this case, growing scholarly attention continues to generate a diverse array of studies addressing the cosmological and astrophysical consequences of this theory \cite{FLMG14, FLMG15, FLMG16, FLMG17, FLMG18, FLMG19, FLMG20, FLMG21, FLMG22}, with more recent works extending its applicability across different regimes of gravitational physics \cite{FLMG23, FLMG24, FLMG25, FLMG26}. 

The DM candidate  plays a central role in both astrophysical phenomena and in the exploration of exotic spacetime geometries, particularly wormholes, which within GR usually demand exotic matter violating classical energy bounds such as the Null Energy Condition (NEC) \cite{DM2, DM3, DM4}. To overcome this limitation, alternative approaches based on quantum gravity have been considered, with Loop Quantum Gravity (LQG) and its cosmological reduction, Loop Quantum Cosmology (LQC), offering important corrections in the high-density regime through the introduction of a critical density $\rho_c$, above which quantum effects modify the classical description of spacetime \cite{DM5}. Within such frameworks, wormhole configurations were successfully constructed without requiring exotic matter, thanks to quantum geometric contributions \cite{DM6, DM7}. Also, this perspective naturally opens the possibility of employing DM as a supporting source for wormholes, given its gravitationally attractive but non-luminous nature. In this context, cosmological measurements testing that DM constitutes nearly five-sixths of the total matter in the universe \cite{DM8}, with its existence strongly confirmed by astrophysical probes including galactic rotation curves, lensing signatures, and cluster dynamics \cite{DM9, DM10, DM11, DM12}. Despite such compelling evidence, the fundamental nature of DM remains elusive, with candidate models ranging from ultralight bosonic fields on the order of $10^{-22}$ eV to heavy Weakly Interacting Massive Particles (WIMPs), all beyond the Standard Model of particle physics \cite{DM13}. In this case, several hypotheses remain under investigation: primordial black holes (BHs) \cite{B1,B2,B3,B4,B5,B6,B7,B8,B9,B10} produced in the early universe may account for part of the DM content, while particles such as axions or WIMPs could annihilate within celestial objects like the Sun, producing neutrinos observable by large detectors such as IceCube, though no conclusive signal has been reported \cite{DM14, DM15}. More recently, innovative detection methods include proposals for solar-orbit satellites to constrain DM properties in the local environment \cite{DM16, DM17}. In addition to these astrophysical searches, DM has been theoretically investigated as a viable wormhole source in both Einsteinian and modified gravitational settings, where its isotropic or anisotropic pressures can satisfy wormhole flare-out conditions without invoking unphysical exotic fluids \cite{DM18, DM19, DM20, DM21, DM22, DM23, DM24, DM25}. Another important analysis tested the construction of traversable wormholes supported by isotropic DM within the LQC framework. In this case, the study of DM density models, \textbf{(i)} the Navarro-Frenk-White (NFW), \textbf{(ii)} the Pseudo-Isothermal (PI), and \textbf{(iii)} the Perfect Fluid (PF) distributions, each offering distinct insights into the behavior of galactic halos and their role in sustaining wormhole geometries under quantum gravitational corrections \cite{DM26, DM27}.

The King model is a statistical mechanics framework originally developed to describe globular clusters but later extended to DM halos, providing a realistic way to avoid the infinite mass problem of classical isothermal cases. It introduces a truncation in the distribution function by assuming that particles with energies above a certain escape energy leave the system, leading to a self-consistent density profile with a flat core and a finite radius. Also, this structure naturally combines an isothermal core and halo with a polytropic envelope of index $n = 5/2$, producing density profiles that decrease as $r^{-3}$ at large distances, consistent with observed galactic halos such as those described by the Burkert profile. Because of evaporation and collisions, the King model captures the evolution of self-gravitating systems toward marginal stability, explaining why many globular clusters and large DM halos are found close to this equilibrium state \cite{King}.

The Dekel-Zhao (DZ) DM density profile has gained prominence as one of the most adaptable models for describing the distribution of DM halos across astrophysical systems \cite{DZM1, DZM2}. In this context, formulated as a generalized double power-law structure, it provides a smooth interpolation between the central and outer regions of galaxies through a set of adjustable parameters that govern the inner slope, the steepness of the transition, and the asymptotic outer behavior \cite{DZM3, DZM4}. In this case, the DZ profile is capable of reproducing a wide range of observed galactic density configurations, making it a robust tool for both theoretical modeling and observational studies in galactic dynamics. Beyond its astrophysical applications, it has also been tested in gravitational contexts, where, testing the influence of a parameter, the profile can give rise to regular or singular BH geometries within the Schwarzschild framework \cite{DZM5, DZM6, DZM7}. Its utility extends further when incorporated into approaches such as the generalized Einstein cluster method, which tests the impact of DM on compact objects in strong gravitational regimes \cite{DZM8, DZM9, DZM10}. In this context, other measurements have illustrated the influence of the DZ model: Khatri et al. \cite{DZM11} demonstrated that the DZ profile can support stable wormhole geometries in Einstein gravity with minimal exotic matter, accompanied by distinctive gravitational lensing effects, while Errehymy et al. \cite{DZM12} showed that traversable wormholes can also emerge in $f(R, L_m,T)$ gravity, where coupling constants mediate matter-geometry interactions. 

Our study tests traversable wormhole geometries in the framework of $f(R,L_{m})$ gravity, the fundamental equations of $f(R, L_{m})$ gravity, and derives the corresponding Einstein field equations tailored for wormhole spacetimes. Also, the analysis of energy conditions is then carried out, providing the essential constraints on the exotic matter content required to sustain the wormhole. Next, we examine Model I, defined by the functional form $f(R, L_{m})=(R/2)+L_{m}^{\alpha}$, where wormhole solutions are constructed and studied under the King and Dekel-Zhao DM density profiles. We are also testing Model II, based on $f(R, L_{m})=(R/2)+(1+\lambda R)L_{m}$, where the wormhole solutions are obtained for the same DM models, offering a comparative perspective between the two cases. Also, the embedding diagrams of the wormhole spacetime are analyzed to visualize the geometry, alongside the evaluation of tidal forces and the computation of the total gravitational energy, which illustrate more information about the wormhole’s stability and traversability. Another important result is that the study also considers the strong deflection angle, emphasizing its relevance for gravitational lensing phenomena and the potential for observational signatures. In this context, we discuss the results in these models, show their physical influence, and broader implications for wormhole physics in the context of modified gravity theories and DM models. 

The paper is organized as follows: Section-\ref{sec1} introduces the wormhole geometries in the framework of $f(R, L_m)$ gravity. Also, in Section-\ref{sec2}, we present the fundamental equations governing this modified gravity theory. In Section-\ref{sec3}, we mention the traversability criteria for wormhole, derives the corresponding Einstein field equations for wormhole spacetime, and analyze the energy conditions, providing the necessary constraints on the matter content supporting the wormhole. Section-\ref{sec4} explores Model-I, where the functional form of the theory is taken as $f(R, L_{m})=\left(R/2\right)+L_{m}^{\alpha}$, and wormhole solutions are obtained in the context of both the King and Dekel–Zhao DM density profiles, discussed separately in subsections-\ref{sec4a} and \ref{sec4b}. In Section-\ref{sec5}, we examine Model-II, with $f(R, L_{m})=\left(R/2\right)+(1+\lambda R)L_{m}$, again analyzing wormhole solutions for the King and Dekel–Zhao models. Section-\ref{sec6} addresses the embedding surface of the wormhole geometry and computes the total gravitational energy. The strong deflection angle and its implications for gravitational lensing are discussed in Section-\ref{sec7}. In this context, Section-\ref{sec8} illustrates and explains the result, showing its physical influence.

%%%%%%%%%%%%%%%%%%%%%%%%%%%%%%%%%%%%%%%%%%%%%%%%%%%%%%%%%%%%%%%%%%%%%%%%%
\section{Basic equations in $f(R, L_m)$ gravity}\label{sec2}
%%%%%%%%%%%%%%%%%%%%%%%%%%%%%%%%%%%%%%%%%%%%%%%%%%%%%%%%%%%%%%%%%%%%
In modified gravity theories, a key approach is the generalization of the action, providing a more flexible framework to describe gravitational dynamics beyond general relativity and enabling the inclusion of additional curvature or matter terms to model cosmic acceleration, DM effects, and other phenomena. In this context, Harko et al. \cite{th10} proposed the $f(R, L_m)$ gravity theory, which generalizes the $f(R)$ models by assuming that the gravitational Lagrangian can be represented as an arbitrary function of the Ricci scalar $R$ and the matter Lagrangian $L_m$. The action corresponding to this framework is expressed as
 \begin{eqnarray}
     \mathcal{S} = \int f(R,L_m)\sqrt{-g}d^4x,\label{action}
 \end{eqnarray}
where $g$ denotes the determinant of the metric tensor $g_{\mu\nu}$. The Ricci scalar curvature $R$ can be obtained by contracting the Ricci tensor $R_{\mu\nu}$ as
 \begin{eqnarray}
     R = g^{\mu\nu}R_{\mu\nu},\label{R}
 \end{eqnarray}
where the Ricci tensor $R_{\mu\nu}$  can be defined as 
\begin{eqnarray}
    R_{\mu\nu} = \partial_\lambda\Gamma^\lambda_{\mu\nu}-\partial_\nu\Gamma^\lambda_{\mu\mu}+\Gamma^\sigma_{\mu\mu}\Gamma^\lambda_{\sigma\lambda}-\Gamma^\lambda_{\mu\sigma}\Gamma^\sigma_{\mu\lambda}.
\end{eqnarray}

Here, $\Gamma^\alpha_{\beta\gamma}$ denotes the components of the Levi-Civita connection, expressed as
\begin{eqnarray}
  \Gamma^\alpha_{\beta\gamma} = \frac{1}{2}g^{\alpha\lambda}\left[\frac{\partial g_{\gamma\lambda}}{\partial x^\beta}+\frac{\partial g_{\lambda\beta}}{\partial x^\gamma}-\frac{\partial g_{\beta\gamma}}{\partial x^\lambda}\right].  
\end{eqnarray}

%%%%%%%%%%%%%%%%%%%%%%%%%%%%%%%%%%%%%%%%%%%%%%%%%%%%%%%%%%%%%%%%%%%%%
\begin{figure}[h]
\begin{center}
\begin{tabular}{rl}
\includegraphics[width=5.7cm]{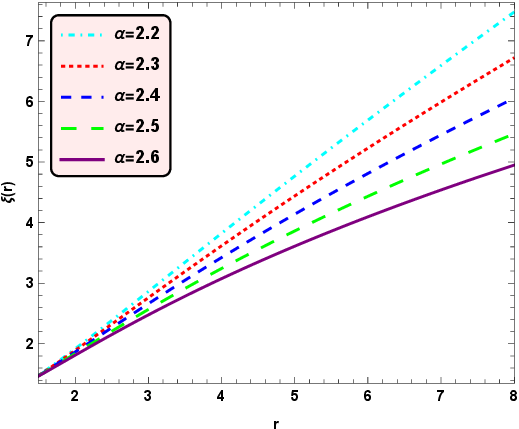}
\includegraphics[width=5.7cm]{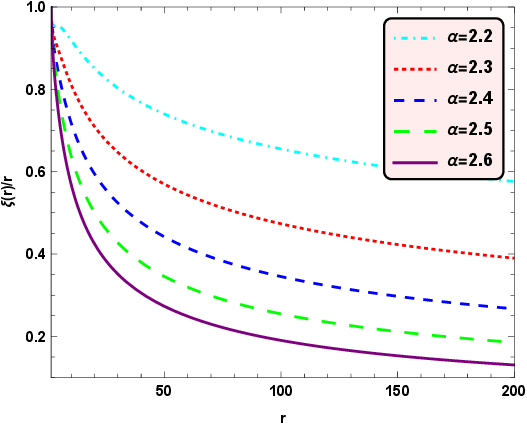}
\includegraphics[width=5.7cm]{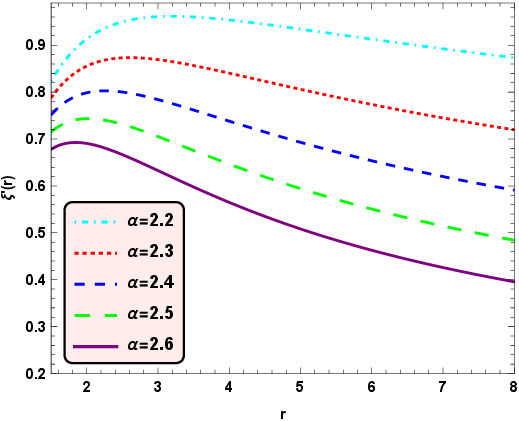}
\\
\end{tabular}
\end{center}
\caption{\label{fig1} Shows the characteristics of shape function $\xi(r)$ (Left), the ratio $\xi(r)/r$ (Middle), and the derivative $\xi'(r)$ (Right) against the radial coordinate $r$ for King DM model under the  $f(R, L_m)$ gravity model-I with parameters $\beta= 0.65$, $\gamma = 1$, $\eta = -0.5$, $r_s$ = 1.01, and $r_0 = 1.4$.}
\end{figure}
%%%%%%%%%%%%%%%%%%%%%%%%%%%%%%%%%%%%%%%%%%%%%%%%%%%%%%%%%%

The field equations of $f(R, L_m)$ gravity follow from the variation of the general action (\ref{action}) with respect to the metric tensor $g_{\mu\nu}$, yielding
\begin{eqnarray}
   f_RR_{\mu\nu} + \left(g_{\mu\nu}\square-\nabla_\mu\nabla_\nu\right)f_R-\frac{1}{2}\left(f-f_{L_m}L_m\right)g_{\mu\nu} = \frac{1}{2}f_{L_m}T_{\mu\nu},\label{FE}
\end{eqnarray}
 where $f_R =\frac{\partial f}{\partial R}$, $f_{L_m}=\frac{\partial f}{\partial L_m}$, and $T_{\mu\nu}$ denotes the energy-momentum tensor of the matter distribution, expressed as
 \begin{eqnarray}
     T_{\mu\nu} = \frac{-2}{\sqrt{-g}}\frac{\delta(\sqrt{-g}L_m)}{\delta g^{\mu\nu}} = g_{\mu\nu}L_m - 2\frac{\partial L_m}{\partial g^{\mu\nu}}.
 \end{eqnarray}

Now, the contraction of the field equation~(\ref{FE}) leads to a relation connecting the energy-momentum scalar $T$, the matter Lagrangian $L_m$, and the Ricci scalar $R$ as
 \begin{eqnarray}
     R f_R +3\square f_R-2(f-f_{L_m}L_m) = \frac{1}{2} f_{L_m}T,
 \end{eqnarray}
where $\square$ stands for the d’Alembertian operator, defined as $\square F = \frac{1}{\sqrt{-g}}\partial_\alpha (\sqrt{-g}g^{\alpha\beta}\partial_\beta F)$ for any scalar function $F$.

Further, the covariant divergence of the energy-momentum tensor in this gravity framework can be expressed as
\begin{equation}
\nabla_\mu T^{\mu\nu} = 2\left[\nabla_\mu\ln(f_{L_m})\right]\frac{\partial L_m}{\partial g_{\mu\nu}}.
\end{equation}

The conservation of the matter energy-momentum tensor, $\nabla_\mu T^{\mu\nu} = 0$, imposes a functional constraint between the matter Lagrangian density and the function $f_{L_m}(R, L_m)$ as
\begin{equation}
\nabla_\mu\ln(f_{L_m})\frac{\partial L_m}{\partial g_{\mu\nu}} = 0. \label{conservation}
\end{equation}
Consequently, given a specific matter Lagrangian density, an appropriate choice of the function $f(R, L_m)$ can yield conservative models with arbitrary matter-geometry coupling.

%%%%%%%%%%%%%%%%%%%%%%%%%%%%%%%%%%%%%%%%%%%%%%%%%%%%%%%%%%%%%%%%%%%%%
\begin{figure}[h]
	\begin{center}
		\begin{tabular}{rl}
			\includegraphics[width=5.7cm]{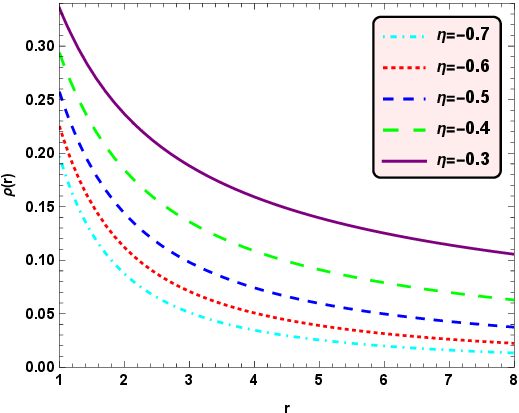}
			\includegraphics[width=5.7cm]{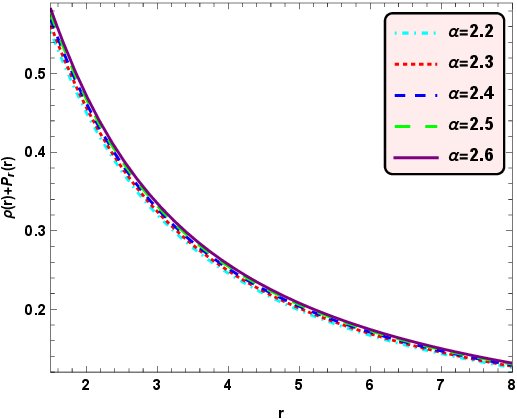}
			\includegraphics[width=5.7cm]{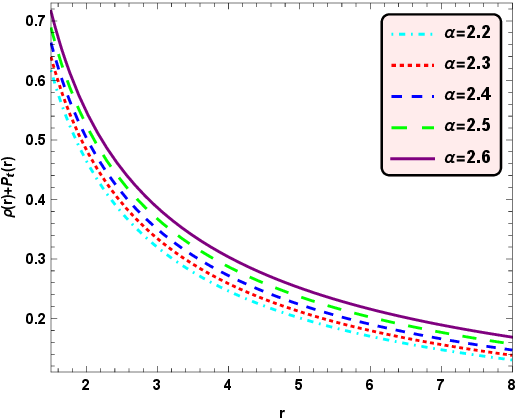}
			\\
		\end{tabular}
		\begin{tabular}{rl}
			\includegraphics[width=5.6cm]{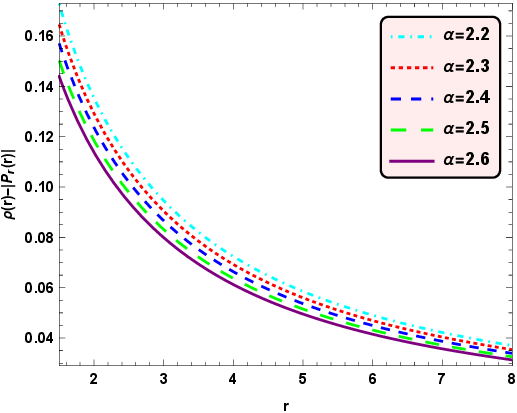}
			\includegraphics[width=5.6cm]{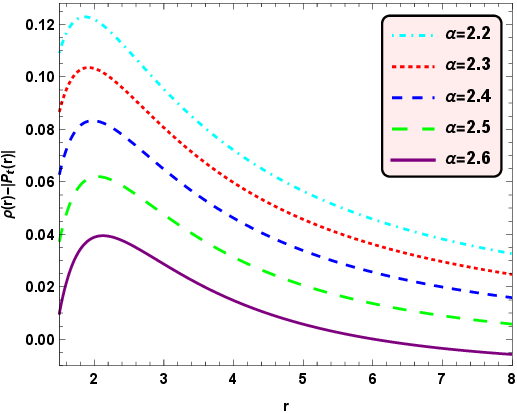}
			\includegraphics[width=5.6cm]{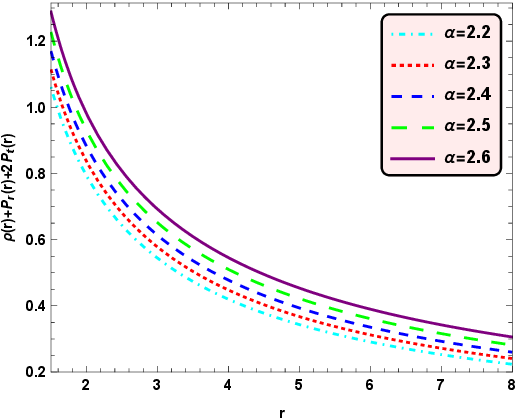}
			\\
		\end{tabular}
	\end{center}
	\caption{ \label{fig2} Shows the characteristic of energy density $\rho(r)$ (Left), $\rho(r)+P_r(r)$ (Middle), $\rho(r)+P_t(r)$ (Right) in the above panel, and  $\rho(r)-|P_r(r)|$ (Left), $\rho(r)-|P_t(r)|$ (Middle), $\rho(r)+P_r(r)+2P_t(r)$ (Right) in the below panel for King DM model under the  $f(R, L_m)$ gravity model-I with parameters $\beta= 0.65$, $\gamma = 1$, $\eta = -0.5$, $r_s$ = 1.01, and $r_0 = 1.4$.}
\end{figure}
%%%%%%%%%%%%%%%%%%%%%%%%%%%%%%%%%%%%%%%%%%%%%%%%%%%%%%%%%%

\section{Field equations for traversable wormhole in $f(R, L_m)$ gravity}\label{sec3}
In this section, we discuss the traversability criteria of the wormhole structure and formulate the Einstein field equations for a traversable wormhole geometry within the framework of $f(R, L_m)$ gravity theory.

\subsection{Traversability Criteria for Wormhole}\label{sec3a}
In this study, we consider a static, spherically symmetric Morris-Thorne wormhole, described by the following line element \cite{MT88}
\begin{equation}
ds^2 = -e^{2\Phi(r)}dt^2+\left(1-\frac{\xi(r)}{r}\right)^{-1}dr^2+r^2(d\theta^2 + sin^2\theta d\phi^2),\label{Metric}
\end{equation}
where $\Phi(r)$ and $\xi(r)$ are the gravitational redshift function and shape function of the wormhole, respectively. The wormhole throat, a crucial structural feature, corresponds to the minimum radius $r = r_0$ for which $\xi(r_0) = r_0$, known as the throat condition. Further, for the wormhole line element (\ref{Metric}),  the proper radial distance function $l(r)$ defined as
\begin{eqnarray}
    l(r) =  \pm \int_{r_0}^{r} \frac{dr}{\sqrt{1-\xi(r)/r}}.\label{l}
\end{eqnarray}
Here, the $\pm$ signs denote the upper and lower regions of the wormhole, which are connected through the throat.

For a wormhole to allow traversal, the above functions are required to meet the following criteria:

\textbf{Gravitational redshift function $\Phi(r)$:} The gravitational redshift function, $\Phi(r)$, must remain finite throughout the spacetime to prevent the formation of an event horizon. If $\Phi(r)$ diverges at any point, it would correspond to an infinite redshift, effectively creating a horizon that would block the passage of signals or travelers.

\textbf{Shape function $\xi(r)$:} The shape function $\xi(r)$ must satisfy the following conditions:  \textit{(i)} For all regions outside the throat, the ratio of the shape function to the radial coordinate must not exceed unity, i.e., $\frac{\xi(r)}{r} \leq 1$ for $r \geq r_0$. \textit{(ii)} The flare-out condition, defined as $\xi^{\prime}(r) < 1$ for $r\geq r_0$, and \textit{(iii)} Asymptotic flatness condition, defined as $\frac{\xi(r)}{r} \rightarrow 0$ as $r\rightarrow \infty$, for asymptotic flatness nature of the wormhole.

\textbf{Proper radial distance function $l(r)$:} This function must remain finite across all radial coordinates $r$. Also, it decreases from the upper universe ($l = \infty$) toward the throat ($l=0$), and then increases from the throat toward the lower universe ($l = -\infty$). Moreover, the proper radial distance $l(r)$ must satisfy $|l(r)| \geq r-r_0$.

%%%%%%%%%%%%%%%%%%%%%%%%%%%%%%%%%%%%%%%%%%%%%%%%%%%%%%%%%%%%%%%%%%%%%
\begin{figure}[h]
	\begin{center}
		\begin{tabular}{rl}
			\includegraphics[width=5.7cm]{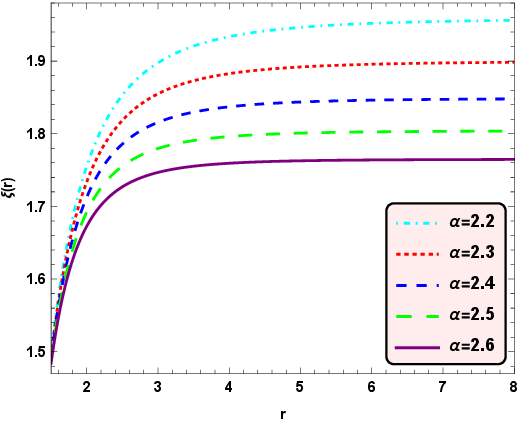}
			\includegraphics[width=5.7cm]{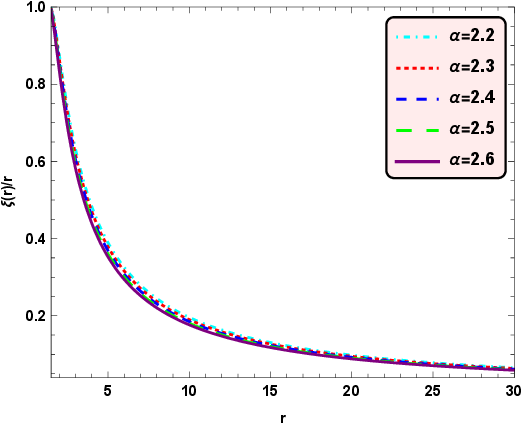}
			\includegraphics[width=5.7cm]{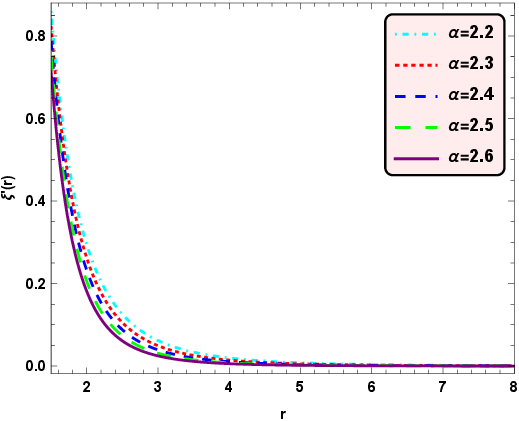}
			\\
		\end{tabular}
	\end{center}
	\caption{\label{fig3} Shows the characteristics of shape function $\xi(r)$ (Left), the ratio $\xi(r)/r$ (Middle), and the derivative $\xi'(r)$ (Right) against the radial coordinate $r$ for Dekel-Zhao DM model under the  $f(R, L_m)$ gravity model-I with parameters $\rho_s= 0.06$, $\kappa = 2.12$, $r_s$ = 6, and $r_0 = 1.4$.}
\end{figure}
%%%%%%%%%%%%%%%%%%%%%%%%%%%%%%%%%%%%%%%%%%%%%%%%%%%%%%%%%%

\subsection{Filed Equations for Traversable Wormhole}\label{sec3b}
Here, we derive the Einstein field equations for the wormhole metric (\ref{Metric}) in $f(R, L_m)$ gravity framework. In this context, the non-vanishing components of the Ricci tensor for the wormhole metric (\ref{Metric}) obtained as
\begin{align}
R_{00} &= e^{2\Phi} \Bigg[ \left(1 - \frac{\xi}{r}\right) \left(\Phi'' + \Phi'^2 + \frac{2\Phi'}{r} \right) - \frac{(r \xi' - \xi)}{2 r^2} \, \Phi' \Bigg], \label{R00} \\
R_{11} &= -\Phi'' - \Phi'^2 + \frac{(r \xi' - \xi)}{2 r (r - \xi)} \left(\Phi' + \frac{2}{r}\right), \label{R11} \\
R_{22} &= (\xi - r)\Phi' + \frac{\xi'}{2} + \frac{\xi}{2r}, \label{R22} \\
R_{33} &= \sin^2 \theta \, \Big[(\xi - r)\Phi' + \frac{\xi'}{2} + \frac{\xi}{2r}\Big]. \label{R33}
\end{align}

Thus, the Ricci curvature scalar $R$ for the wormhole geometry (\ref{Metric}) can be obtained using equation~(\ref{R}) as 
\begin{eqnarray}
    R = \frac{2\xi'}{r^2}-2\left(1-\frac{\xi}{r}\right)\left(\Phi''+\Phi'^2+\frac{\Phi'}{r}\right)+\frac{\Phi'}{r^2}(r\xi'+\xi-2r).
\end{eqnarray}

In this study, the matter distribution of the wormhole structure is assumed to be anisotropic, and the corresponding energy-momentum tensor is expressed
\begin{eqnarray}
    T_{\mu\nu} = [\rho(r)+P_t(r)]u_\mu u_\nu+P_t(r)\delta_{\mu\nu}+[P_r(r)-P_t(r)]u_\mu u_\nu,\label{T}
\end{eqnarray}
where $u^{\mu}$ denotes the four-velocity vector, and $v^{\mu}$ is the unitary space-like vector, satisfying $-u^{\mu}u_{\mu} = v^{\mu}v_{\mu} = 1$. Besides, $\rho(r)$ represents the energy density, while $P_{r}(r)$ and $P_{t}(r)$ represent the radial and tangential pressures, respectively. Notably, they depend solely on the radial coordinate $r$.

Now, the Einstein filed equations (\ref{FE}) for  the wormhole geometry (\ref{Metric}) associated with the anisotropic matter distribution (\ref{T}) read as
\begin{eqnarray}
    && \left(1-\frac{\xi}{r}\right)\left[\left\{ \Phi''+\Phi'^2+\frac{2\Phi'}{r}-\frac{r\xi'-\xi}{2r(r-\xi)}\Phi'\right\}F-\left\{\Phi'+\frac{2}{r}-\frac{r\xi'-\xi}{2r(r-\xi)}\right\}F'-F''\right]+\frac{f-L_mf_{Lm}}{2}= \frac{f_{L_m}}{2}\rho(r),\label{rho}
    \\
    && \left(1-\frac{\xi}{r}\right)\left[\left\{ -\Phi''-\Phi'^2+\frac{r\xi'-\xi}{2r(r-\xi)}\left(\Phi'+\frac{2}{r}\right)\right\}F+\left\{\Phi'+\frac{2}{r}-\frac{r\xi'-\xi}{2r(r-\xi)}\right\}F'\right]-\frac{f-L_mf_{Lm}}{2} = \frac{f_{L_m}}{2}P_r(r),\label{pr}
    \\
    && \left(1-\frac{\xi}{r}\right)\left[\left\{-\frac{\Phi'}{r}+\frac{r\xi'+\xi}{2r^2(r-\xi)}\right\}F+\left\{\Phi'+\frac{2}{r}-\frac{r\xi'-\xi}{2r(r-\xi)}\right\}F'+F''\right]-\frac{f-L_mf_{Lm}}{2} = \frac{f_{L_m}}{2}P_t(r).\label{pt}
\end{eqnarray}
Here, $F = \frac{\partial f}{\partial R}$. It is important to note that the above field equations provide a foundation for exploring a wide range of traversable wormhole solutions within the framework of $f(R, L_{m})$ gravity.

%%%%%%%%%%%%%%%%%%%%%%%%%%%%%%%%%%%%%%%%%%%%%%%%%%%%%%%%%%%%%%%%%%%%%
\begin{figure}[h]
	\begin{center}
		\begin{tabular}{rl}
			\includegraphics[width=5.7cm]{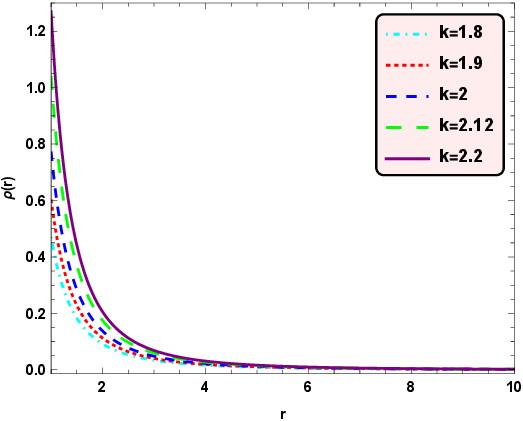}
			\includegraphics[width=5.7cm]{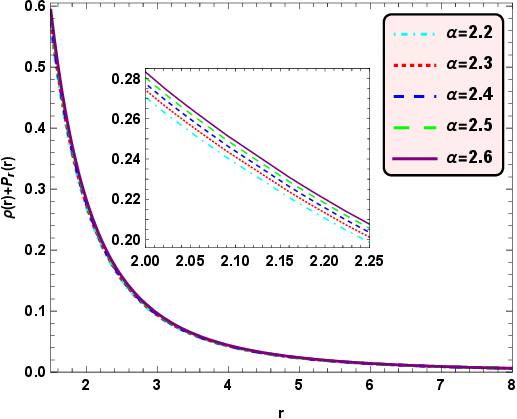}
			\includegraphics[width=5.7cm]{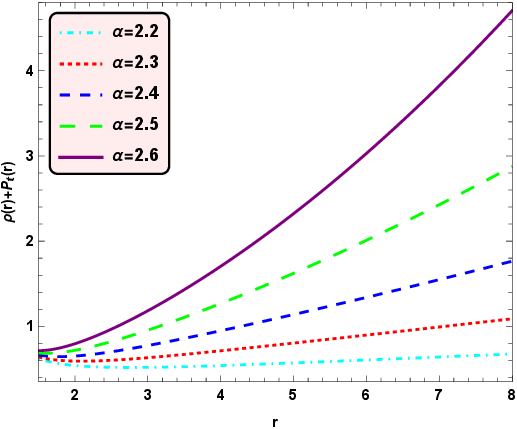}
			\\
		\end{tabular}
		\begin{tabular}{rl}
			\includegraphics[width=5.7cm]{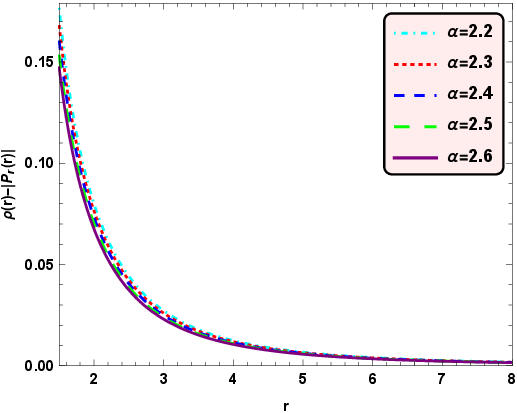}
			\includegraphics[width=5.7cm]{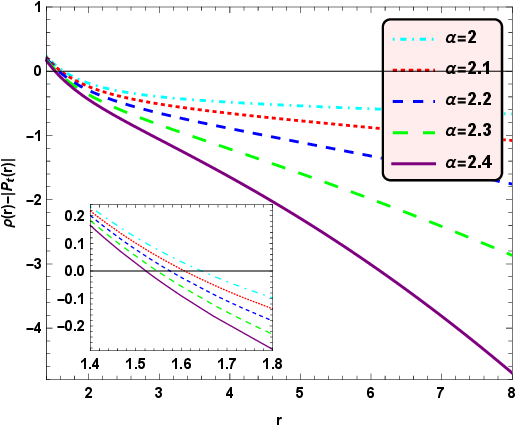}
			\includegraphics[width=5.7cm]{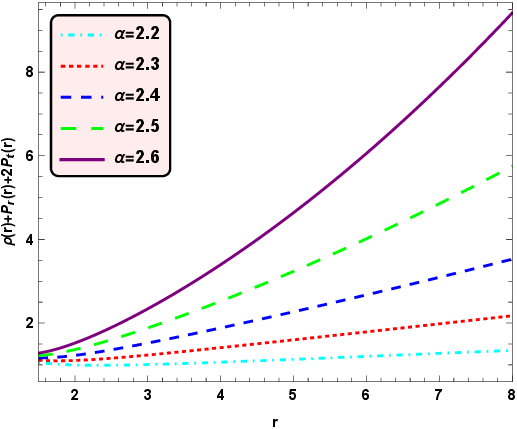}
			\\
		\end{tabular}
	\end{center}
	\caption{\label{fig4} Shows the characteristic of energy density $\rho(r)$ (Left), $\rho(r)+P_r(r)$ (Middle), $\rho(r)+P_t(r)$ (Right) in the above panel, and  $\rho(r)-|P_r(r)|$ (Left), $\rho(r)-|P_t(r)|$ (Middle), $\rho(r)+P_r(r)+2P_t(r)$ (Right) in the below panel for Dekel-Zhao DM model under the  $f(R, L_m)$ gravity model-I with parameters $\rho_s= 0.06$, $\kappa = 2.12$, $r_s$ = 6.5, and $r_0 = 1.4$.}
\end{figure}
%%%%%%%%%%%%%%%%%%%%%%%%%%%%%%%%%%%%%%%%%%%%%%%%%%%%%%%%%%

We now briefly address the classical energy conditions, derived from the Raychaudhuri equations. In the framework of general relativity, the traversable wormholes demand exotic matter by violating the null energy condition (NEC) \cite{MT88, ms88a}. In contrast, modified gravity theories can comply with or avoid this condition because their effective field equations are not identical to those in general relativity. The Raychaudhuri equations, describing the temporal evolution of congruences of timelike vectors $u^\eta$ and the null geodesics $k_\eta$, can be expressed as \cite{ar55}
\begin{eqnarray}
     &&\frac{d\Theta}{d\tau}-\omega_{\eta\xi}\omega^{\eta\xi}+\sigma_{\eta\xi}\sigma^{\eta\xi}+\frac{1}{3}\theta^2+R_{\eta\xi}u^\eta u_\xi = 0,
     \\
     &&\frac{d\Theta}{d\tau}-\omega_{\eta\xi}\omega^{\eta\xi}+\sigma_{\eta\xi}\sigma^{\eta\xi}+\frac{1}{3}\theta^2+R_{\eta\xi}k^\eta k_\xi = 0,
\end{eqnarray}
where $k^\eta$ represents the vector files, $R_{\eta\xi}k^\eta k_\xi$ is the shear or spatial tensor with $\sigma ^2 = \sigma_{\eta\xi}\sigma^{\eta\xi} \geq 0$ and $\omega_{\eta\xi}\equiv 0$. In the scenario of attractive gravity ($\theta < 0$), the Raychaudhuri equations indicate the following conditions
\begin{eqnarray}
   R_{\eta\xi}u^\eta u_\xi \geq 0,~~ R_{\eta\xi}k^\eta k_\xi \geq 0.
\end{eqnarray}

Thus, the energy conditions for anisotropic matter take the form:

\textit{(i)} Null Energy Condition (NEC):
\begin{align*}
\text{NEC}_r: \quad & \rho(r) + P_r(r) \geq 0, \\
\text{NEC}_t: \quad & \rho(r) + P_t(r) \geq 0.
\end{align*}

\textit{(ii)} Weak energy condition (WEC): 
\begin{align*}
\text{WEC}_r: \quad & \rho(r) \geq 0,~\rho(r) + P_r(r) \geq 0,\\
\text{WEC}_t: \quad & \rho(r) \geq 0,~ \rho(r) + P_t(r) \geq 0.
\end{align*}

\textit{(ii)} Dominant energy condition (DEC): 
\begin{align*}
\text{DEC}_r: \quad & \rho(r) \geq 0,~\rho(r) - |P_r(r)| \geq 0,\\
\text{DEC}_t: \quad & \rho(r) \geq 0,~ \rho(r) - |P_t(r)| \geq 0.
\end{align*}

\textit{(iv)} Strong energy condition (SEC): $\rho(r) + P_r(r) \geq 0, ~ \rho(r) + P_t(r) \geq 0, ~ \rho(r) + P_r(r)+2P_t(r) \geq 0$.

To construct physically viable wormhole solutions within the framework of \( f(R, \mathcal{L}_m) \) gravity, we consider two specific and well-motivated functional forms of the gravitational Lagrangian, model-I: $f(R, L_m) = \frac{R}{2} + L_m^{\alpha}$, and model-II: $f(R, L_m) = \frac{R}{2} + (1 + \lambda R)L_m$. These forms are chosen to capture different modes of curvature-matter coupling and to facilitate analytical tractability of the field equations. 

%%%%%%%%%%%%%%%%%%%%%%%%%%%%%%%%%%%%%%%%%%%%%%%%%%%%%%%%%%%%%%%%%%%%%
\begin{figure}[h]
\begin{center}
\begin{tabular}{rl}
\includegraphics[width=4.25cm]{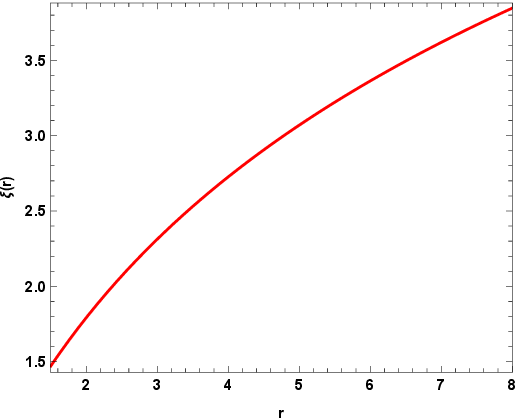}
\includegraphics[width=4.25cm]{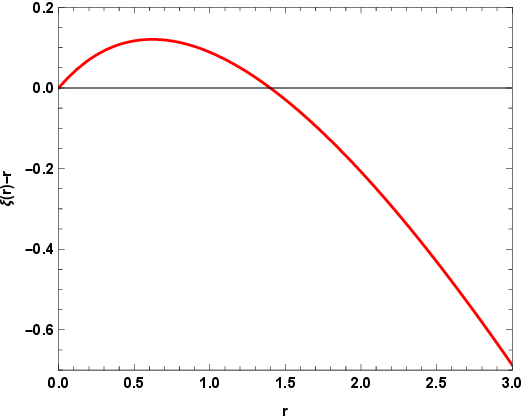}
\includegraphics[width=4.25cm]{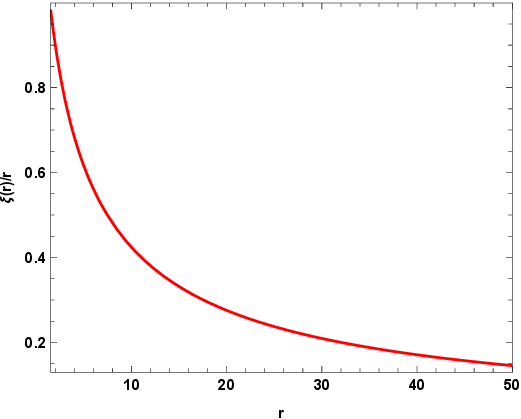}
\includegraphics[width=4.25cm]{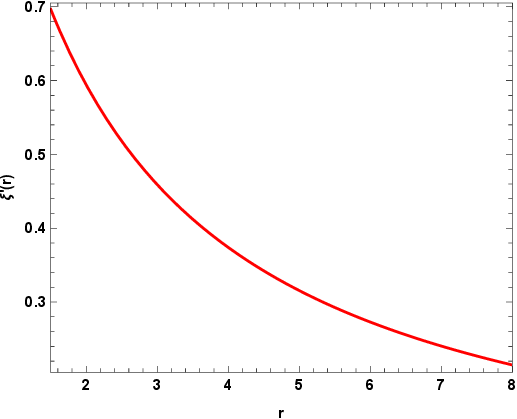}
\\
\end{tabular}
\end{center}
\caption{\label{fig5} Shows the characteristics of wormhole shape function $\xi(r)$ (\ref{B3})(Left), the function $\xi(r) - r$ (Left-Middle), the ratio $\xi(r)/r$ (Right-Middle), and the derivative $\xi'(r)$ (Right) against the radial coordinate $r$ under the  $f(R, L_m)$ gravity model-II with parameter $r_0 = 1.4$.}
\end{figure}
%%%%%%%%%%%%%%%%%%%%%%%%%%%%%%%%%%%%%%%%%%%%%%%%%%%%%%%%%%

%%%%%%%%%%%%%%%%%%%%%%%%%%%%%%%%%%%%%%%%%%%%%%%%%%%%%%%%%%%%%%%%%%%
\section{Model-I: Wormhole Solutions for $f(R, L_m)$=$\frac{R}{2}+L_m^\alpha$}\label{sec4}
%%%%%%%%%%%%%%%%%%%%%%%%%%%%%%%%%%%%%%%%%%%%%%%%%%%%%%%%%%%%%%%%%%%
In this section, we derive the wormhole solutions by considering the following minimal functional form of $f(R, L_m)$\cite{lv22}
\begin{eqnarray}
    f(R, L_m)=\frac{R}{2}+L_m^\alpha.\label{f1}
\end{eqnarray}

Here, the parameter $\alpha$ serves as a crucial free variable, whose value can be adjusted to modulate the physical properties of the system. Remarkably, for $\alpha = 1$, the model naturally reproduces the well-known wormhole geometry within the framework of general relativity (GR). Indeed, this particular choice simplifies the underlying field equations while retaining the essential coupling between the matter Lagrangian density $L_m$  and the Ricci scalar $R$. Such a minimal form enables the exploration of the fundamental geometrical and physical characteristics of wormhole configurations without invoking additional complexities arising from higher-order or non-minimal curvature-matter interactions.

For a traversable wormhole, the tidal forces experienced by a traveler must remain within human tolerance, not exceeding Earth’s surface gravity $g_\oplus$. Focusing on radial motion in the equatorial plane ($\theta = \pi/2$), the condition $|\Delta {\bf a}| \leq g_\oplus$ ensures safe passage, yielding the following constraint on the radial tidal force at the wormhole throat \cite{MT88}
\begin{eqnarray}
    |\Phi'(r_0)|&\leq& \frac{g_\oplus r_0}{1-\xi'(r_0)},\label{dphi}
\end{eqnarray}
Since the constant redshift function satisfies the above constraint, we adopt it in this study.

Thus, for a constant redshift function and with the choice $L_{m} = -\rho(r)$, the Einstein field equations (\ref{rho})–(\ref{pt}) in this model simplify to the following form 
\begin{eqnarray}
    \rho(r) &=& \left(\frac{\xi'(r)}{(2\alpha-1)r^2}\right)^{\frac{1}{\alpha}},\label{rho1}
    \\
    P_r(r) &=& \frac{\rho(r)\left[(\alpha-1)r^3-\xi(r) (\rho(r))^\alpha \right]}{\alpha r^3},\label{pr1}
    \\
    P_t(r) &=& \frac{(\rho(r))^{1-\alpha}\left[\xi(r)-r\xi'(r)+2(\alpha-1)r^3 (\rho(r))^\alpha \right]}{2\alpha r^3}.\label{pt1}
\end{eqnarray}

The above system of field equations comprises three equations with four unknowns. Hence, to obtain viable wormhole solutions within DM halos, we adopt two specific DM halo models: the King DM model and the Dekel–Zhao DM model.

%%%%%%%%%%%%%%%%%%%%%%%%%%%%%%%%%%%%%%%%%%%%%%%%%%%%%
\subsection{King Dark Matter Model}\label{sec4a}
%%%%%%%%%%%%%%%%%%%%%%%%%%%%%%%%%%%%%%%%%%%%%%%%%%%%%%

The King DM density profile is a widely used phenomenological model for describing galactic DM distributions \cite{ir72}. Owing to its consistency with low surface brightness and flat rotation curves, it effectively models core-like DM halos \cite{an18}. The King DM density profile is defined as \cite{ir72, an18}
\begin{equation}
\rho(r)=\beta\left[\left(\frac{r}{r_s}\right)^2+\gamma\right]^\eta,\label{KDM}
\end{equation}
where $\beta$, $\gamma$, $\eta$ are parameters, and $r_s$ denotes the
 scale radius. 

Substituting the above energy density (\ref{KDM}) into the field equation (\ref{rho1}), the corresponding shape function $\xi(r)$ is obtained as
\begin{eqnarray}
    \xi(r)= \mathcal{C}_1+\frac{1}{3} (2 \alpha -1) r^3 \beta ^{\alpha } \gamma ^{\alpha  \eta } F(r),
\end{eqnarray}
where $F(r)={_2F_1}\left(\frac{3}{2},-\alpha  \eta ;\frac{5}{2};-\frac{r^2}{\gamma r_s^2 }\right)$ is the hypergeometric function, and $\mathcal{C}_1$ is an integration constant. Here, the throat condition $\xi(r_0) = r_0$ is applied to determine the constant $\mathcal{C}_1$, which yields the following expression for $\mathcal{C}_1$
\begin{eqnarray}
    \mathcal{C}_1 = r_0-\frac{1}{3} (2 \alpha -1) r_0^3 \beta ^{\alpha } \gamma ^{\alpha  \eta } F(r_0).
\end{eqnarray}

Thus, with the above values of $\mathcal{C}_1$, the resulting shape function takes the following form
\begin{eqnarray}
\xi(r)&=& r_0+\frac{1}{3} (2 \alpha -1)\beta ^{\alpha } \gamma ^{\alpha  \eta } \,\left[r^3F(r)-r_0^3 F(r_0)\right].\label{B1}
\end{eqnarray}

Furthermore, the wormhole solutions are required to satisfy the flare-out condition at the throat, given by the relation
\begin{eqnarray}
\xi'(r_0)&=& (2 \alpha -1) r_0^2 \beta ^{\alpha } \gamma ^{\alpha  \eta } \left(1+\frac{r_0^2}{\gamma  r_s^2}\right)^{\alpha \eta } < 1. \label{b1}
\end{eqnarray}

The obtained shape function (\ref{B1}) is illustrated graphically in Fig. \ref{fig1} for the $f(R, L_m)$ gravity parameter values $\alpha = 2.2, 2.3, 2.4, 2.5, 2.6$ with $\beta= 0.65$, $\gamma = 1$, $\eta = -0.5$, $r_s$ = 1.01, and $r_0 = 1.4$.  Fig. \ref{fig1} shows that the shape function $\xi(r)$ increases monotonically with the radial coordinate $r$ and decreases with increasing parameter $\alpha$, while consistently satisfying $\xi(r)/r \leq 1$ for $r \geq r_0$. Moreover, the shape function meets the flare-out condition in this configuration. These results ensure that the shape function (\ref{B1}) derived from the King DM density profile (\ref{KDM}) successfully captures all essential features of a traversable wormhole. Notably, the shape function also exhibits the desired asymptotic flat behavior, ensuring that the resulting traversable wormhole geometries are asymptotically flat.

The radial and transverse pressures are then obtained from the field equations (\ref{pr})-(\ref{pt}), resulting in the following expressions
\begin{eqnarray}
    P_r(r) &=& \frac{\beta}{3\alpha  r^3}  \left(\gamma +\frac{r^2}{r_s^2}\right)^{\eta } \left[ \beta^\alpha  \left(\gamma +\frac{r^2}{r_s^2}\right)^{\eta\alpha} \left\{(2 \alpha -1)\beta ^{\alpha } \gamma ^{\eta \alpha } (r_0^3F(r_0)-r^3F(r))-3r_0\right\}+3(\alpha -1) r^3\right],
    \\
    P_t(r) &=& \frac{\beta ^{1-\alpha }}{6 \alpha  r^3}\left(\gamma +\frac{r^2}{r_s^2}\right)^{(1-\alpha ) \eta } \left[(2 \alpha -1) \beta ^{\alpha } \gamma ^{\eta \alpha } \left(r^3 F(r) -r_0^3F(r_0)\right)+3r_0-3 r^3 \beta ^{\alpha } \gamma ^{\eta \alpha } \left(1+\frac{r^2}{\gamma  r_s^2}\right)^{\alpha  \eta }\right].
\end{eqnarray}

%%%%%%%%%%%%%%%%%%%%%%%%%%%%%%%%%%%%%%%%%%%%%%%%%%%%%%%%%%%%%%%%%%%%
\begin{figure}[h]
	\begin{center}
		\begin{tabular}{rl}
			\includegraphics[width=5.7cm]{rho11.eps}
			\includegraphics[width=5.7cm]{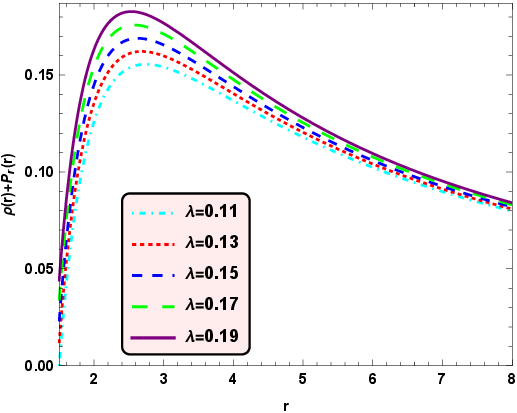}
			\includegraphics[width=5.7cm]{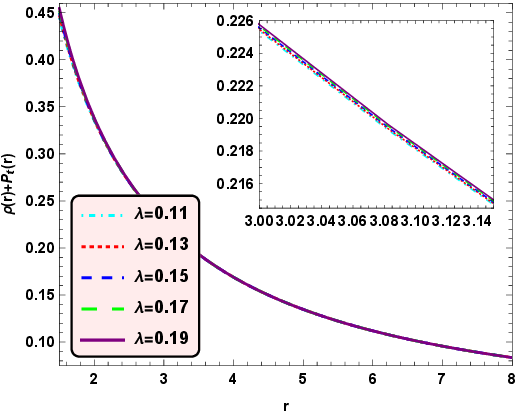}
			\\
		\end{tabular}
		\begin{tabular}{rl}
			\includegraphics[width=5.7cm]{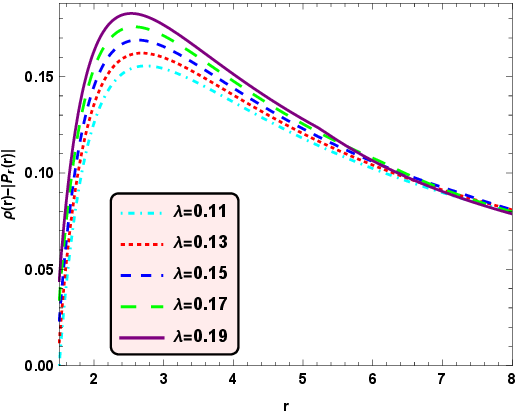}
			\includegraphics[width=5.7cm]{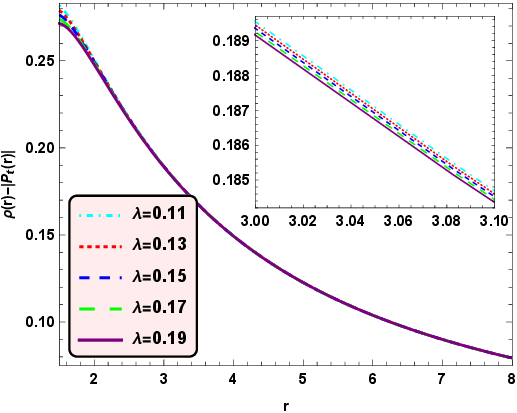}
			\includegraphics[width=5.7cm]{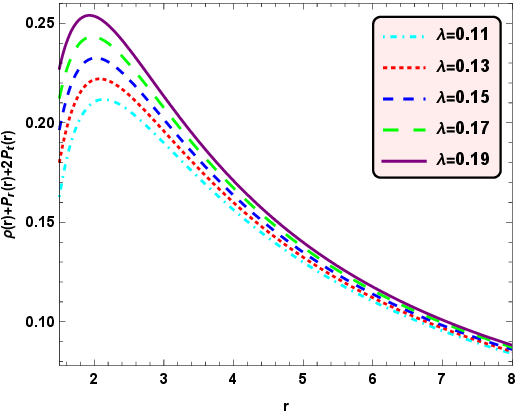}
			\\
		\end{tabular}
	\end{center}
	\caption{\label{fig6} Shows the characteristic of energy density $\rho(r)$ (Left), $\rho(r)+P_r(r)$ (Middle), $\rho(r)+P_t(r)$ (Right) in the above panel, and  $\rho(r)-|P_r(r)|$ (Left), $\rho(r)-|P_t(r)|$ (Middle), $\rho(r)+P_r(r)+2P_t(r)$ (Right) in the below panel for King DM model under the  $f(R, L_m)$ gravity model-II with parameters $\beta= 0.65$, $\gamma = 1$, $\eta = -0.5$, $r_s$ = 1.01, and $r_0 = 1.4$.}
\end{figure}
%%%%%%%%%%%%%%%%%%%%%%%%%%%%%%%%%%%%%%%%%%%%%%%%%%%%%%

In this model, the NEC, DEC, and SEC at the wormhole throat can be explicitly expressed as
\begin{eqnarray}
\left[\rho(r)+P_{r}(r)\right]_{r=r_{0}}&=& \frac{\beta}{\alpha r_0^2}\left(\gamma +\frac{r_0^2}{r_s^2}\right)^{\eta } \left[(2 \alpha -1)r_0^2-\beta ^{\alpha } \left(\gamma +\frac{r_0^2}{r_s^2}\right)^{\alpha  \eta }\right] > 0, 
\\
\left[\rho(r)+P_{t}(r)\right]_{r=r_{0}}&=& \frac{\beta ^{1-\alpha }}{2 \alpha r_0^2}\left(\gamma +\frac{r_0^2}{r_s^2}\right)^{(1-\alpha ) \eta } \left[1+(2 \alpha -1) d^2 \beta ^{\alpha } \left(\gamma +\frac{r_0^2}{r_s^2}\right)^{\alpha  \eta }\right] > 0,
\\
\left[\rho(r)-|P_{r}(r)|\right]_{r=r_{0}}&=& \beta  \left(\gamma +\frac{r_0^2}{r_s^2}\right)^{\eta }-\left|\frac{\beta}{\alpha r_0^2}  \left(\gamma+\frac{r_0^2}{r_s^2} \right)^{\eta } \left[(\alpha -1)r_0^2- \beta ^{\alpha }\left(\gamma +\frac{r_0^2}{r_s^2}\right)^{\alpha \eta }\right]\right| > 0,
\\
\left[\rho(r)-|P_{t}(r)|\right]_{r=r_{0}}&=& \beta  \left(\gamma +\frac{r_0^2}{r_s^2}\right)^{\eta }-\left|\frac{3 \beta ^{1-\alpha }}{\alpha  r_0^2} \left(\gamma +\frac{r_0^2}{r_s^2}\right)^{(1-\alpha) \eta} \left[1-r_0^2 \beta ^{\alpha } \left(\gamma +\frac{r_0^2}{r_s^2}\right)^{\alpha\eta}\right]\right| > 0,
\\
\left[\rho(r)+P_{r}(r)+2P_{t}(r)\right]_{r=r_{0}}&=& \frac{\beta^{1-\alpha }}{\alpha  r_0^2}\left(\gamma +\frac{r_0^2}{r_s^2}\right)^{(1-\alpha ) \eta } \left[1-\beta ^{2 \alpha } \left(\gamma +\frac{r_0^2}{r_s^2}\right)^{2\alpha\eta }+2 (\alpha -1)r_0^2 \beta^{\alpha} \left(\gamma +\frac{r_0^2}{r_s^2}\right)^{\alpha\eta }\right] > 0.\nonumber
\\
\end{eqnarray}

To investigate the physical nature of the matter content supporting the wormholes, we provide graphical illustrations of all relevant energy conditions in Fig.~\ref{fig2} corresponding to the parameters $\alpha = 2.2, 2.3, 2.4, 2.5, 2.6$ with $\beta = 0.65$, $\gamma = 1$, $\eta = -0.5$, $r_s = 1.01$, and $r_0 = 1.4$. Notably, the energy conditions NEC$_r$, NEC$_t$, WEC$_r$, WEC$_t$, DEC$_r$, and SEC are satisfied throughout the wormhole spacetime, while DEC$_t$ is satisfied near the throat, indicating the absence of exotic matter. These findings confirm that the wormhole solutions can be sustained within the King DM halo under $f(R, L_m)$ gravity without requiring exotic matter.

%%%%%%%%%%%%%%%%%%%%%%%%%%%%%%%%%%%%%%%%%%%%%%%%%%%%%%%%%%%%%%%%%%%%
\begin{figure}[h]
	\begin{center}
		\begin{tabular}{rl}
			\includegraphics[width=5.7cm]{rho12.eps}
			\includegraphics[width=5.7cm]{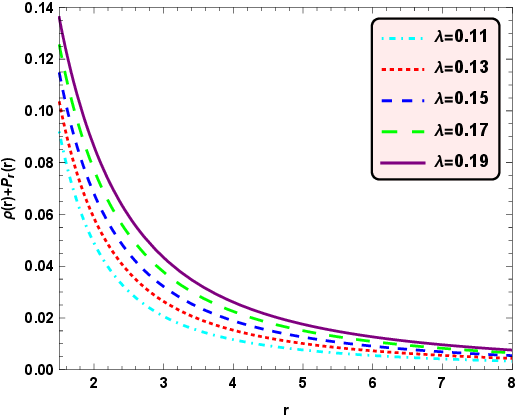}
			\includegraphics[width=5.7cm]{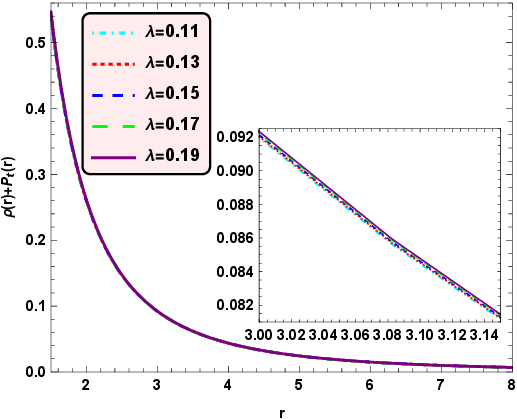}
			\\
		\end{tabular}
		\begin{tabular}{rl}
			\includegraphics[width=5.7cm]{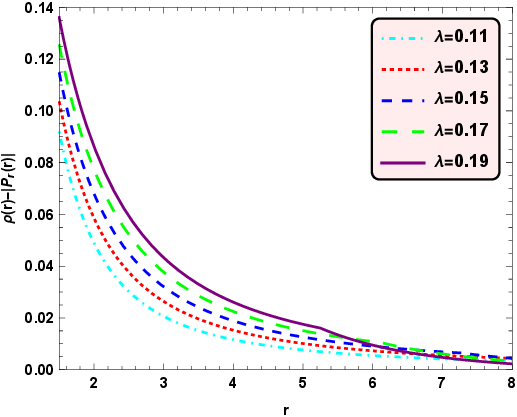}
			\includegraphics[width=5.7cm]{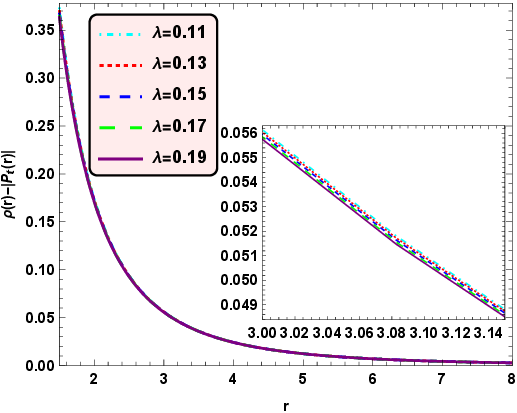}
			\includegraphics[width=5.7cm]{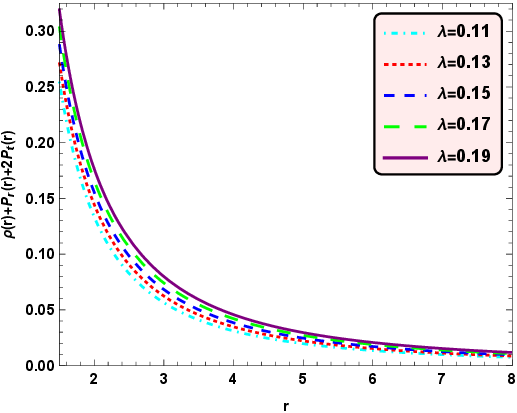}
			\\
		\end{tabular}
	\end{center}
	\caption{\label{fig7} Shows the characteristic of energy density $\rho(r)$ (Left), $\rho(r)+P_r(r)$ (Middle), $\rho(r)+P_t(r)$ (Right) in the above panel, and  $\rho(r)-|P_r(r)|$ (Left), $\rho(r)-|P_t(r)|$ (Middle), $\rho(r)+P_r(r)+2P_t(r)$ (Right) in the below panel for Dekel-Zhao DM model under the  $f(R, L_m)$ gravity model-II with parameters $\rho_s= 0.06$, $\kappa = 2.12$, $r_s$ = 6.5, and $r_0 = 1.4$.}
\end{figure}
%%%%%%%%%%%%%%%%%%%%%%%%%%%%%%%%%%%%%%%%%%%%%%%%%%%%%%

%%%%%%%%%%%%%%%%%%%%%%%%%%%%%%%%%%%%%%%%%%%%%%%%%%%%%%%%%%%%%%%%%%%
\subsection{Dekel-Zhao Dark Matter Model}\label{sec4b}
%%%%%%%%%%%%%%%%%%%%%%%%%%%%%%%%%%%%%%%%%%%%%%%%%%%%%%%%%%%%%%%%%%%
In this section, we adopt the Dekel–Zhao (DZ) DM density profile to construct traversable wormholes. The DZ profile is a widely used, versatile model for DM halos, capable of capturing a broad range of astrophysical structures \cite{DZ1, DZ2}. Its double power-law form smoothly interpolates between inner and outer slopes, making it well-suited for both observational and theoretical studies. The profile is defined as \cite{DZ1, DZ2,DZ4}
\begin{equation}
\rho(r)=\frac{\rho_s}{\left(\frac{r}{r_s}\right)^\kappa\left[1+\left(\frac{r}{r_s}\right)^{1/b}\right]^{b(d-\kappa)}}.\label{dz}
\end{equation}
Here, the parameters $\rho_s$ and $r_s$ represent the characteristic density and the scale radius, respectively. Moreover, the parameters $\kappa$, $b$, and $d$ control the inner slope, the transition sharpness, and the outer slope of the density profile. For $b = N_1$ and $d = 3 + N_2/N_1$ with natural numbers $N_1$ and $N_2$, simplified analytical expressions for the gravitational potential, enclosed mass, and velocity dispersion can be obtained. In this study, we choose $N_1 = 2$ and $N_2 = 1$, so the DZ density profile (\ref{dz}) becomes
\begin{equation}
\rho(r)=\frac{\rho_s}{\left(\frac{r}{r_s}\right)^\kappa\left[1+\left(\frac{r}{r_s}\right)^{1/2}\right]^{7-2\kappa}},\label{DZ}
\end{equation}

In this case, the shape function is obtained by substituting the above density profile (\ref{DZ}) into the field equation (\ref{rho1}), giving:
\begin{eqnarray}
    \xi(r)&=& \mathcal{C}_2+\frac{2 \alpha -1}{3-\alpha\kappa} r^3 \rho_s^{\alpha } \left(\frac{r_s}{r}\right)^{\alpha\kappa}H(r),
\end{eqnarray}
where $H(r)={_2F_1}\left[\alpha  (7-2 k),6-2 \alpha  k,7-2 \alpha  k,-\sqrt{\frac{r}{s}}\right]$, and $\mathcal{C}_2$ is an integration constant. Here, the throat condition $\xi(r_0) = r_0$ yields the following result
\begin{eqnarray}
    \mathcal{C}_2 = r_0+\frac{2 \alpha -1}{\alpha  \kappa-3} r_0^3 \rho_s^{\alpha } \left(\frac{r_s}{r_0}\right)^{\alpha  \kappa}H(r_0).
\end{eqnarray}

Consequently, incorporating the above result of $\mathcal{C}_2$, the shape function can be expressed in its final form as
\begin{eqnarray}
\xi(r)&=& r_0+\frac{2 \alpha -1}{3-\alpha\kappa}\rho_s^{\alpha }\left[ r^3 \left(\frac{r_s}{r}\right)^{\alpha\kappa}H(r)- r_0^3 \left(\frac{r_s}{r_0}\right)^{\alpha\kappa}H(r_0)\right].\label{B2}
\end{eqnarray}

In this case, the flare-out condition at the wormhole throat is given by the following relation
\begin{eqnarray}
\xi'(r_0)&=&  (2\alpha -1) r_0^2 r\rho_s^{\alpha }  \left(\frac{r_s}{r_0}\right)^{\alpha  \kappa}\left(1+\sqrt{\frac{r_0}{r_s}}\right)^{\alpha(2 \kappa-7)} < 1. \label{b1}
\end{eqnarray}

The derived shape function (\ref{B2}) is shown in Fig.~\ref{fig3} for $\alpha = 2.2, 2.3, 2.4, 2.5, 2.6$ with parameters $\rho_s = 0.06$, $\kappa = 2.12$, $r_s = 6$, and $r_0 = 1.4$. As illustrated, the shape function increases monotonically with $r$, satisfies $\xi(r)/r \leq 1$ for $r \geq r_0$, and meets the flare-out condition. Moreover, $\xi(r)/r \rightarrow 0$ as $r \rightarrow \infty$, demonstrating the desired asymptotic behavior. Therefore, the shape function derived from the DZ DM density profile (\ref{rho2}) successfully satisfies the essential features of asymptotically flat traversable wormhole geometries.

%%%%%%%%%%%%%%%%%%%%%%%%%%%%%%%%%%%%%%%%%%%%%%%%%%%%%%

\begin{figure}[h]
	\begin{center}
		\begin{tabular}{rl}
			\includegraphics[width=5.7cm]{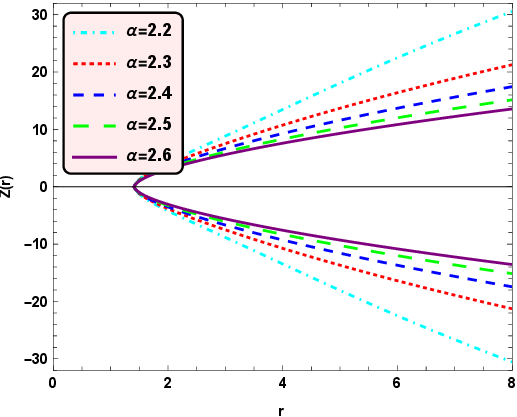}
			\includegraphics[width=5.7cm]{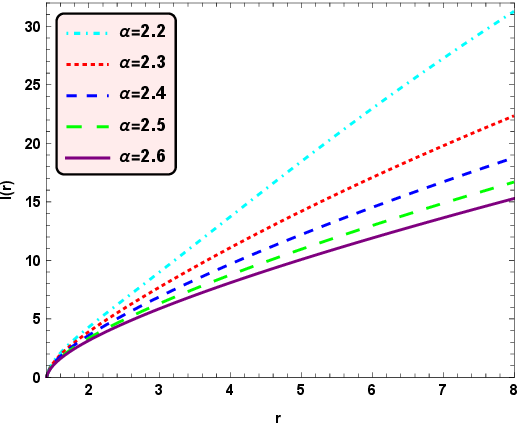}
			\includegraphics[width=5.3cm]{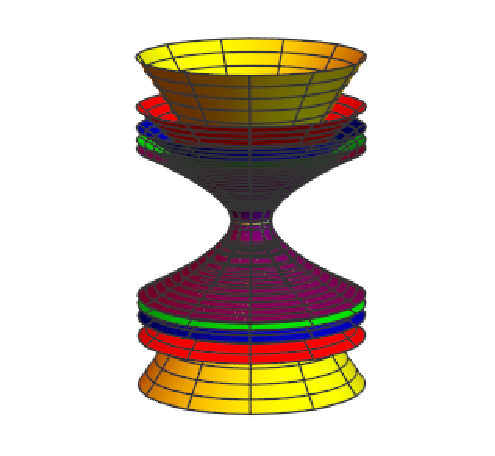}
			\\
		\end{tabular}
	\end{center}
	\caption{\label{fig8} Shows the characteristics of embedding surface (Left), proper radial length (Middle) against the radial coordinate $r$, and the full visualization diagram of wormholes (Right) for King DM model under the  $f(R, L_m)$ gravity model-I with parameters $\beta= 0.65$, $\gamma = 1$, $\eta = -0.5$, $r_s$ = 1.01, and $r_0 = 1.4$. In the full visualization diagram of wormholes: $\alpha = 2.2$ (Yellow), $\alpha = 2.3$ (Red), $\alpha = 2.4$ (Blue), $\alpha = 2.5$ (Green), $\alpha = 2.6$ (Purple). }
\end{figure}
%%%%%%%%%%%%%%%%%%%%%%%%%%%%%%%%%%%%%%%%%%%%%%%%%%%%%%

The radial and transverse pressures in this case are obtained as follows
\begin{eqnarray}
    P_r(r) &=& \frac{\rho_s}{\alpha  r^3}\left(\frac{r_s}{r}\right)^\kappa\left(1+\sqrt{\frac{r}{r_s}}\right)^{2\kappa-7}  \Bigg[(\alpha -1) r^3-\frac{\rho_s^{\alpha }}{\alpha  \kappa-3}\left(\frac{r_s}{r}\right)^{\alpha  \kappa} \left(1+\sqrt{\frac{r}{r_s}}\right)^{\alpha  (2 \kappa-7)}  \Bigg\{(2 \alpha -1) \rho_s^{\alpha } \Bigg(r_0^3\left(\frac{r_s}{r_0}\right)^{\alpha  \kappa}\times\nonumber
    \\
    &&H(r_0)-r^3\left(\frac{r_s}{r}\right)^{\alpha  \kappa}H(r)\Bigg)+r_0(\alpha  \kappa-3)\Bigg\}\Bigg].
    \\
    P_t(r) &=& \frac{\rho_s^{1-\alpha } \left(\frac{r_s}{r}\right)^{(1-\alpha ) \kappa}}{2 \alpha  r^3 (\alpha  \kappa-3)} \left(1+\sqrt{\frac{r}{s}}\right)^{(1-\alpha ) (2 \kappa-7)} \Bigg[(\alpha  \kappa-3) \left\{r_0-r^3 \rho_s^{\alpha } \left(\frac{r_s}{r}\right)^{\alpha  \kappa} \left(1+\sqrt{\frac{r}{s}}\right)^{\alpha  (2 \kappa-7)}\right\}+(2\alpha -1)\times\nonumber
    \\
    && \rho_s^{\alpha }\left\{r_0^3\left(\frac{r_s}{r_0}\right)^{\alpha  \kappa}H(r_0)-r^3 \left(\frac{r_s}{r}H(r)\right)^{\alpha \kappa}\right\}\Bigg].
\end{eqnarray}

In this case, at the wormhole throat, the NEC, DEC, and SEC can be expressed as
\begin{eqnarray}
\left[\rho(r)+P_{r}(r)\right]_{r=r_{0}}&=& \frac{\rho_s\mathcal{J}_1}{\alpha  r_0^2} \left[(2\alpha -1)r_0^2-\rho_s^{\alpha } \left(\frac{r_s}{r_0}\right)^{\alpha  k} \left(1+\sqrt{\frac{r_0}{r_s}}\right)^{\alpha  (2\kappa-7)}\right] > 0, 
\\
\left[\rho(r)+P_{t}(r)\right]_{r=r_{0}}&=& \frac{\rho_s^{1-\alpha }\mathcal{J}_1^{1-\alpha }}{2 \alpha  r_0^2}  \left[1+(2\alpha -1)r_0^2 \rho_s^{\alpha }\left(\frac{r_s}{r_0}\right)^{\alpha  \kappa} \left(1+\sqrt{\frac{r_0}{r_s}}\right)^{\alpha(2 \kappa-7)}\right] > 0,
\\
\left[\rho(r)-|P_{r}(r)|\right]_{r=r_{0}}&=& \rho_s\mathcal{J}_1-\left| \frac{\rho_s\mathcal{J}_1}{\alpha r_0^2} \left[(\alpha -1)r_0^2-\rho_s^{\alpha }\left(\frac{r_s}{r_0}\right)^{\alpha  \kappa} \left(1+\sqrt{\frac{r_0}{r_s}}\right)^{\alpha  (2 \kappa-7)}\right]\right|> 0,
\\
\left[\rho(r)-|P_{t}(r)|\right]_{r=r_{0}}&=& \rho_s\mathcal{J}_1-\frac{1}{2} \left| \frac{\rho_s^{1-\alpha }\mathcal{J}_1^{1-\alpha }}{\alpha r_0^2} \left[1-r_0^2 \rho_s^{\alpha } \left(\frac{r_s}{r_0}\right)^{\alpha\kappa}\left(1+\sqrt{\frac{r_0}{r_s}}\right)^{\alpha(2 \kappa-7) }\right]\right| > 0,
\\
\left[\rho(r)+P_{r}(r)+2P_{t}(r)\right]_{r=r_{0}}&=& \frac{\rho_s^{1-\alpha }\mathcal{J}_1^{1-\alpha }}{\alpha  r_0^2} \left[1-\rho_s^{2\alpha }\mathcal{J}_1^{2\alpha }+2 (\alpha -1)r_0^2\rho_s^{\alpha } \left(\frac{r_s}{r_0}\right)^{\alpha  \kappa} \left(1+\sqrt{\frac{r_0}{r_s}}\right)^{\alpha  (2 \kappa-7)}\right] > 0,
\end{eqnarray}
where
\begin{eqnarray}
    \mathcal{J}_1 =\left(\frac{r_s}{r_0}\right)^\kappa \left(1+\sqrt{\frac{r_0}{r_s}}\right)^{2 \kappa-7}.
\end{eqnarray}

Figure~\ref{fig4} illustrates the relevant energy conditions, showing that NEC, WEC, DEC$_r$, and SEC are satisfied throughout the wormhole spacetime, while DEC$_t$ is satisfied near the throat. This indicates the absence of exotic matter and confirms that the wormhole solutions can be supported within the DZ DM halo under $f(R, L_m)$ gravity without requiring exotic matter.

%%%%%%%%%%%%%%%%%%%%%%%%%%%%%%%%%%%%%%%%%%%%%%%%%%%%%%%%%%%%%%%%
\begin{figure}[h]
	\begin{center}
		\begin{tabular}{rl}
			\includegraphics[width=5.6cm]{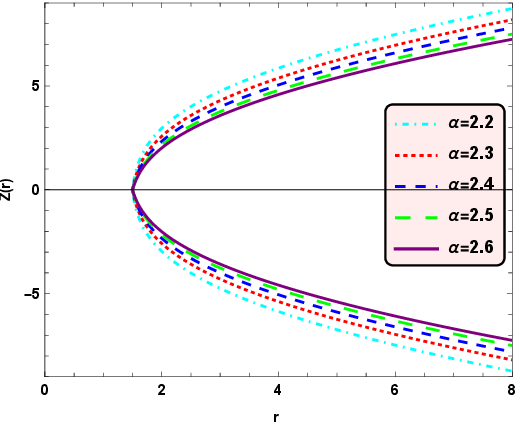}
			\includegraphics[width=5.6cm]{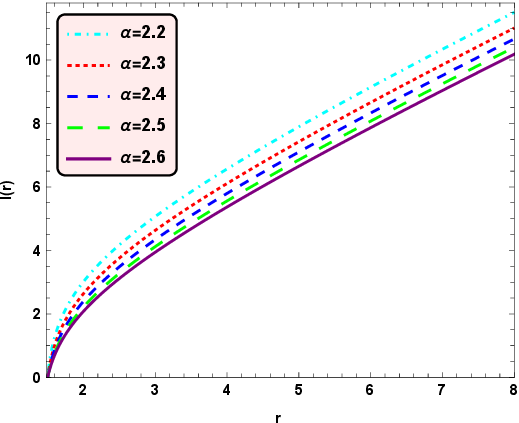}
			\includegraphics[width=5.6cm]{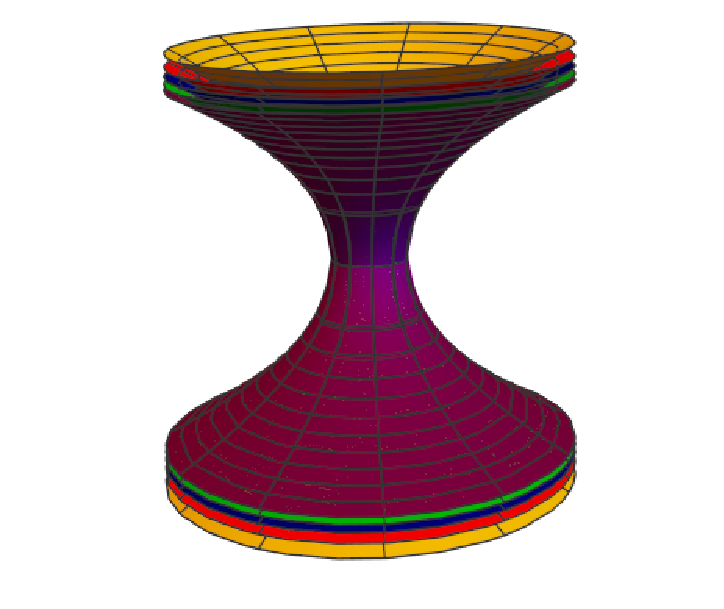}
			\\
		\end{tabular}
	\end{center}
	\caption{\label{fig9} Shows the characteristics of embedding surface (Left), proper radial length (Middle) against the radial coordinate $r$, and the full visualization diagram of wormholes (Right) for Dekel-Zhao DM model under the  $f(R, L_m)$ gravity model-I with parameters $\rho_s= 0.06$, $\kappa = 2.12$, $r_s$ = 6.5, and $r_0 = 1.4$. In the full visualization diagram of wormholes: $\alpha = 2.2$ (Yellow), $\alpha = 2.3$ (Red), $\alpha = 2.4$ (Blue), $\alpha = 2.5$ (Green), $\alpha = 2.6$ (Purple).}
\end{figure}
%%%%%%%%%%%%%%%%%%%%%%%%%%%%%%%%%%%%%%%%%%%%%%%%%%%%%%

%%%%%%%%%%%%%%%%%%%%%%%%%%%%%%%%%%%%%%%%%%%%%%%%%%%%%%%%%%%%%%%%%%%
\section{Model-II: Wormhole Solutions for $f(R, L_m)$=$\frac{R}{2}+\left(1+\lambda R\right)L_m$}\label{sec5}
%%%%%%%%%%%%%%%%%%%%%%%%%%%%%%%%%%%%%%%%%%%%%%%%%%%%%%%%%%%%%%%%%%%

In this section, we derive the wormhole solutions by considering the following non-minimal form of the $f(R, L_m)$ function \cite{nm10, nm11, rv21}
\begin{equation}
f(R,L_m) = \frac{R}{2} + (1 + \lambda R)L_m, \label{fRm2}
\end{equation}
where $\lambda$ denotes the coupling constant. It is noteworthy that for $\lambda = 0$, the model reduces to the standard wormhole geometry of GR.
 
In this model, taking a constant redshift function and $L_{m} = \rho(r)$, the Einstein field equations (\ref{rho})–(\ref{pt}) reduce to the following form  
\begin{align}
\rho(r) &= \frac{\xi'(r)}{r^{2} + 2\lambda \xi'(r)}, \label{rho2}
\\
P_{r}(r) &= \frac{2\lambda r (r^{3} - \xi'(r)) \xi'(r) - (r^{2} + 4\lambda \xi'(r)) \xi(r)}{r \, (r^{2} + 2\lambda \xi'(r))^{2}}, \label{pr2} 
\\
P_{t}(r) &= \frac{r^{2} \xi(r) - (r^{3} - 4\lambda \xi(r))\xi'(r)}{2r \, (r^{2} + 2\lambda \xi'(r))^{2}}. \label{pt2}
\end{align}
As obtaining an exact solution for this non-linear model is challenging, we introduce a new shape function to generate the wormhole solutions under this framework, given by
\begin{eqnarray}
    \xi(r) = \frac{r_0}{\log (2)} \log \left(1+\frac{r}{r_0}\right).\label{B3}
\end{eqnarray}

The shape function (\ref{B3}) is depicted in Fig.~\ref{fig5}, which shows that $\xi(r)$ increases monotonically with the radial coordinate $r$, while $\xi(r) - r$ intersects the $r$-axis at $r = r_0 = 1.4$. It consistently satisfies the condition $\xi(r)/r \leq 1$ for $r \geq r_0$ and the flare-out condition. Therefore, the proposed shape function (\ref{B3}) supports the traversable wormhole geometries by satisfying all the essential characteristics. Moreover, it exhibits the asymptotic flatness, confirming that the resulting traversable wormhole geometries are asymptotically flat.

In this case, the radial and transverse pressures are determined as follows
\begin{eqnarray}
    P_r(r) &=& \frac{2\lambda r_0 r \left[r^3 \log (2) (r_0+r)-r_0\right]-r_0(r+r_0) \left[4\lambda r_0+\log (2) (r+r_0)\right]r^2\log \left(\frac{r+r_0}{r_0}\right)}{r \left[2\lambda r_0 +\log (2) (r+r_0)r^2\right]^2},\label{prII}
    \\
    P_t(r) &=& \frac{r_0(r+r_0) \left[\log \left(\frac{r+r_0}{r_0}\right) \left(4\lambda r_0+ \log (2) (r+r_0)r^2\right)-\log (2)r^3 \right]}{2 r \left[2\lambda r_0  +\log (2) (r_0+r)r^2\right]^2}.\label{ptII}
\end{eqnarray}

Next, we examine all the relevant energy conditions corresponding to the shape function (\ref{B3}) for both the King and Dekel–Zhao DM models.

%%%%%%%%%%%%%%%%%%%%%%%%%%%%%%%%%%%%%%%%%%%%%%%%%%%%%%%%%%
\begin{figure}[h]
\begin{center}
\begin{tabular}{rl}
\includegraphics[width=5.7cm]{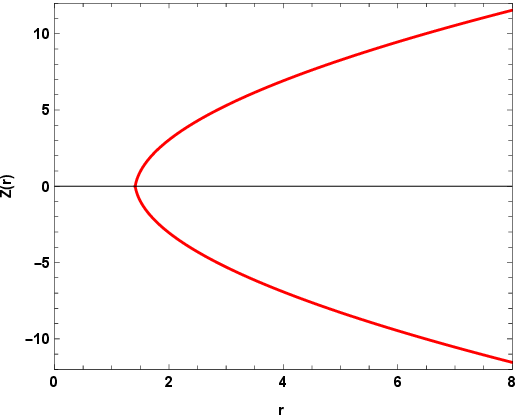}
\includegraphics[width=5.7cm]{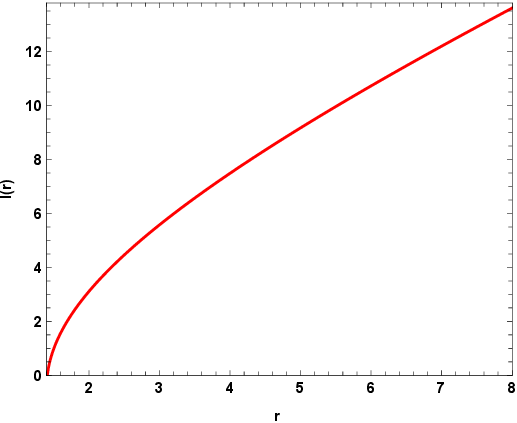}
\includegraphics[width=5.6cm]{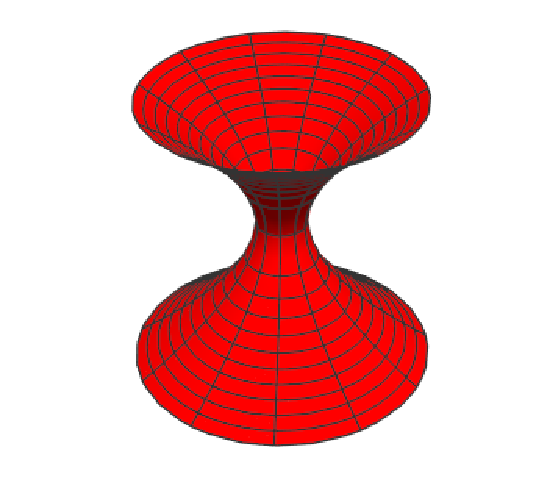}
\\
\end{tabular}
\end{center}
\caption{\label{fig10} Shows the characteristics of embedding surface (Left), proper radial length (Middle) against the radial coordinate $r$, and the full visualization diagram of wormhole (Right) for wormhole shape function (\ref{B3}) under the  $f(R, L_m)$ gravity model-II with parameter $r_0 = 1.4$.}
\end{figure}
%%%%%%%%%%%%%%%%%%%%%%%%%%%%%%%%%%%%%%%%%%%%%%%%%%%%%%%%%%%%%%%%%%%%%

%%%%%%%%%%%%%%%%%%%%%%%%%%%%%%%%%%%%%%%%%%%%%%%%%%%%%
\subsection{King Dark Matter Model}\label{sec5a}
%%%%%%%%%%%%%%%%%%%%%%%%%%%%%%%%%%%%%%%%%%%%%%%%%%%%%%

Here, we analyze the energy density profile of the King DM model (\ref{KDM}) together with the proposed shape function (\ref{B3}) to examine the corresponding energy conditions. This comprehensive analysis will offer valuable insights into the behavior and characteristics of Model~II in sustaining the traversable wormhole structures.

In this model, employing the energy density (\ref{KDM}) along with the radial and transverse pressures (\ref{prII}) and (\ref{ptII}), the NEC, DEC, and SEC at the throat $r = r_0$ are obtained as
\begin{eqnarray}
\left[\rho(r)+P_{r}(r)\right]_{r=r_{0}}&=& \beta  \left(\gamma +\frac{r_0^2}{r_s^2}\right)^{\eta }+\frac{\lambda  \left[2\log (2)(r_0^3-2)-1\right]-2r_0^2 \log ^2(2)}{2 \left[\lambda+r_0^2 \log (2) \right]^2} > 0, 
\\
\left[\rho(r)+P_{t}(r)\right]_{r=r_{0}}&=& \beta  \left(\gamma +\frac{r_0^2}{r_s^2}\right)^{\eta }+ \frac{\log (2) \left[4 \lambda+r_0^2 (2 \log (2)-1) \right]}{4 \left[\lambda+r_0^2 \log (2) \right]^2}> 0,
\\
\left[\rho(r)-|P_{r}(r)|\right]_{r=r_{0}}&=& \beta  \left(\gamma +\frac{r_0^2}{r_s^2}\right)^{\eta }-\left|\frac{\lambda  \left[2\log (2)(r_0^3-2)-1\right]-2r_0^2 \log ^2(2)}{2 \left[\lambda+r_0^2 \log (2) \right]^2}\right|> 0,
\\
\left[\rho(r)-|P_{t}(r)|\right]_{r=r_{0}}&=&\beta  \left(\gamma +\frac{r_0^2}{r_s^2}\right)^{\eta }- \left| \frac{\log (2) \left[4 \lambda+r_0^2 (2 \log (2)-1) \right]}{4 \left[\lambda+r_0^2 \log (2) \right]^2}\right| > 0,
\\
\left[\rho(r)+P_{r}(r)+2P_{t}(r)\right]_{r=r_{0}}&=& \beta  \left(\gamma +\frac{r_0^2}{r_s^2}\right)^{\eta }+\frac{(2\lambda r_0-1)r_0^2 \log (2)-\lambda }{2 \left[\lambda +r_0^2 \log (2)\right]^2} > 0,
\end{eqnarray}

Also, the graphical representation of all relevant energy conditions is shown in Fig.~\ref{fig6} for the parameters $\lambda = 0.11, 0.13, 0.15, 0.17, 0.19$ with $\beta = 0.65$, $\gamma = 1$, $\eta = -0.5$, $r_s = 1.01$, and $r_0 = 1.4$. Remarkably, all energy conditions, NEC, WEC, DEC, and SEC, are satisfied throughout the wormhole spacetime, indicating the absence of exotic matter. These results confirm that the wormhole solutions can be maintained within the King DM halo under $f(R, L_m)$ gravity without the need for exotic matter.

%%%%%%%%%%%%%%%%%%%%%%%%%%%%%%%%%%%%%%%%%%%%%%%%%%%%%%%%%%%%%%%%%%%%%
\begin{figure}[h]
\begin{center}
\begin{tabular}{rl}
\includegraphics[width=4.2cm]{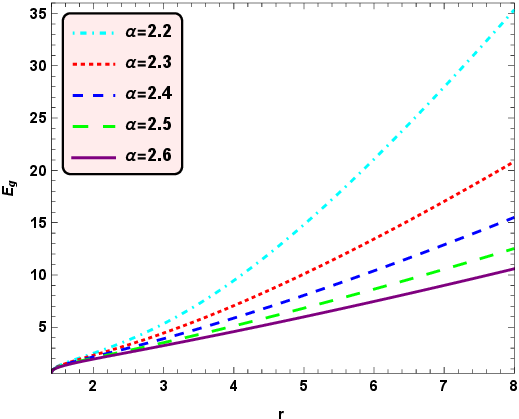}
\includegraphics[width=4.2cm]{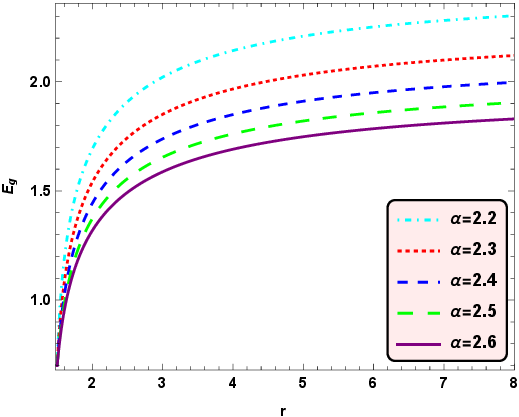}
\includegraphics[width=4.2cm]{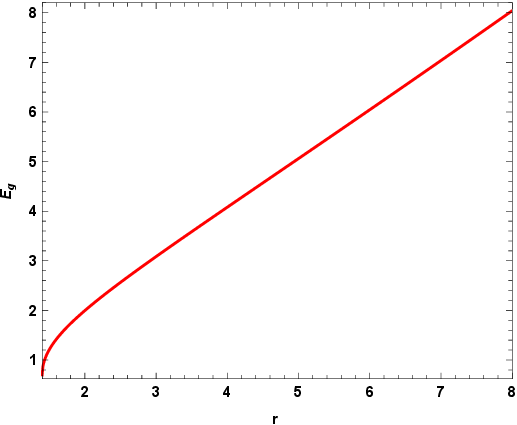}
\includegraphics[width=4.2cm]{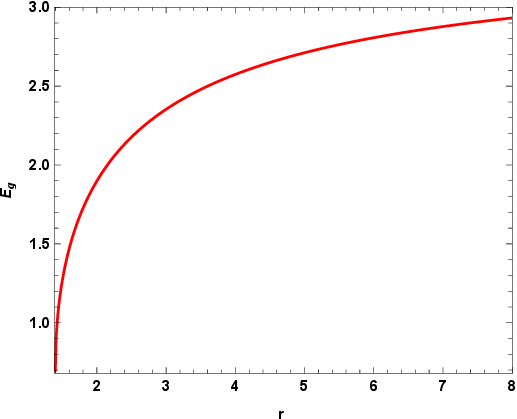}
\\
\end{tabular}
\end{center}
\caption{\label{fig11} Shows the characteristics of total gravitational energy for King DM model under the $f(R, L_m)$ gravity model-I (Left), for Dekel-Zhao DM model under the $f(R, L_m)$ gravity model-I (Left-Middle), for King DM model under the $f(R, L_m)$ gravity model-II (Right-Middle), and for Dekel-Zhao DM model under the $f(R, L_m)$ gravity model-II (Right). In King DM model, parameters $\rho_s= 0.06$, $\kappa = 2.12$, $r_s$ = 6.5, and $r_0 = 1.4$, and in Dekel-Zhao DM model, parameters $\rho_s= 0.06$, $\kappa = 2.12$, $r_s$ = 6.5, and $r_0 = 1.4$. }
\end{figure}
%%%%%%%%%%%%%%%%%%%%%%%%%%%%%%%%%%%%%%%%%%%%%%%%%%%%%%%%%%

%%%%%%%%%%%%%%%%%%%%%%%%%%%%%%%%%%%%%%%%%%%%%%%%%%%%%%%%%%%%%%%%%%%
\subsection{Dekel-Zhao Dark Matter Model}\label{sec5b}
%%%%%%%%%%%%%%%%%%%%%%%%%%%%%%%%%%%%%%%%%%%%%%%%%%%%%%%%%%%%%%%%%%%

In this subsection, we examine the energy conditions associated with the energy density profile (\ref{DZ}) of the DZ DM model and the proposed shape function (\ref{B3}). Specifically, the NEC, DEC, and SEC at the wormhole throat $r = r_0$ are calculated using the energy density (\ref{DZ}) and the radial and transverse pressures (\ref{prII}) and (\ref{ptII}), yielding
\begin{eqnarray}
\left[\rho(r)+P_{r}(r)\right]_{r=r_{0}}&=& \rho_s\left(\frac{r_s}{r_0}\right)^k \left(1+\sqrt{\frac{r_0}{r_s}}\right)^{2 \kappa-7}+\frac{\lambda  \left[2\log (2)(r_0^3-2)-1\right]-2r_0^2 \log ^2(2)}{2 \left[\lambda+r_0^2 \log (2) \right]^2} > 0, 
\\
\left[\rho(r)+P_{t}(r)\right]_{r=r_{0}}&=& \rho_s\left(\frac{r_s}{r_0}\right)^k \left(1+\sqrt{\frac{r_0}{r_s}}\right)^{2 \kappa-7}+ \frac{\log (2) \left[4 \lambda+r_0^2 (2 \log (2)-1) \right]}{4 \left[\lambda+r_0^2 \log (2) \right]^2}> 0,
\\
\left[\rho(r)-|P_{r}(r)|\right]_{r=r_{0}}&=& \rho_s\left(\frac{r_s}{r_0}\right)^k \left(1+\sqrt{\frac{r_0}{r_s}}\right)^{2 \kappa-7}-\left|\frac{\lambda  \left[2\log (2)(r_0^3-2)-1\right]-2r_0^2 \log ^2(2)}{2 \left[\lambda+r_0^2 \log (2) \right]^2}\right|> 0,
\\
\left[\rho(r)-|P_{t}(r)|\right]_{r=r_{0}}&=&\rho_s\left(\frac{r_s}{r_0}\right)^k \left(1+\sqrt{\frac{r_0}{r_s}}\right)^{2 \kappa-7}- \left| \frac{\log (2) \left[4 \lambda+r_0^2 (2 \log (2)-1) \right]}{4 \left[\lambda+r_0^2 \log (2) \right]^2}\right| > 0,
\\
\left[\rho(r)+P_{r}(r)+2P_{t}(r)\right]_{r=r_{0}}&=& \rho_s\left(\frac{r_s}{r_0}\right)^k \left(1+\sqrt{\frac{r_0}{r_s}}\right)^{2 \kappa-7}+\frac{(2\lambda r_0-1)r_0^2 \log (2)-\lambda }{2 \left[\lambda +r_0^2 \log (2)\right]^2} > 0,
\end{eqnarray}

In this model, all the energy conditions, NEC, WEC, DEC, and SEC, are satisfied throughout the wormhole spacetime for parameter $\lambda = 0.11, 0.13, 0.15, 0.17, 0.19$ with $\rho_s= 0.06$, $\kappa = 2.12$, $r_s$ = 6.5, and $r_0 = 1.4$, as clear from Fig. \ref{fig7}. Thus, the wormhole solutions supported by the proposed shape function (\ref{B3}) can also be sustained within the DZ DM halo under $f(R, L_m)$ gravity without exotic matter.

%%%%%%%%%%%%%%%%%%%%%%%%%%%%%%%%%%%%%%%%%%%%%%%%%%%%%%%%%%
\begin{figure}[h]
\begin{center}
\begin{tabular}{rl}
\includegraphics[width=5.7cm]{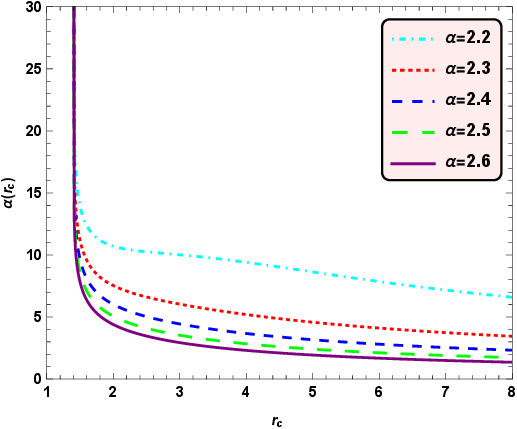}
\includegraphics[width=5.7cm]{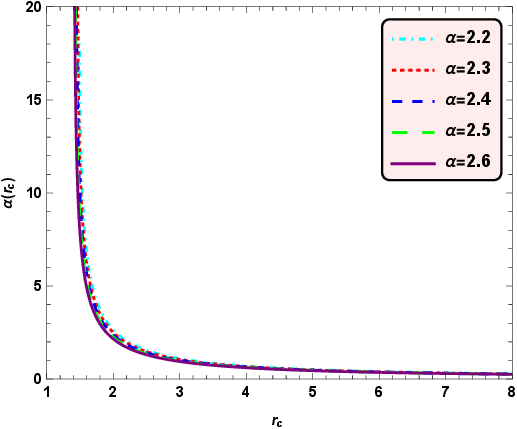}
\includegraphics[width=5.7cm]{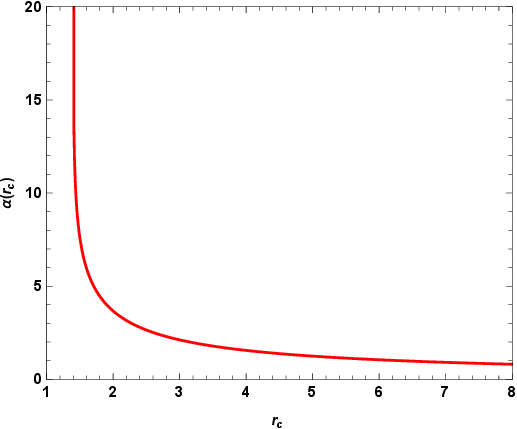}
\\
\end{tabular}
\end{center}
\caption{\label{fig12} Shows the characteristics of deflection angles for King DM model under the $f(R, L_m)$ gravity model-I with parameters $\rho_s= 0.06$, $\kappa = 2.12$, $r_s$ = 6.5, and $r_0 = 1.4$ (Left), for Dekel-Zhao DM model under the $f(R, L_m)$ gravity model-I with parameters $\rho_s= 0.06$, $\kappa = 2.12$, $r_s$ = 6.5, and $r_0 = 1.4$ (Middle), and for wormhole shape function (\ref{B3}) under the $f(R, L_m)$ gravity model-II with parameter $r_0 = 1.4$ (Right).}
\end{figure}
%%%%%%%%%%%%%%%%%%%%%%%%%%%%%%%%%%%%%%%%%%%%%%%%%%%%%%%%%%%%%%%%%%%%%

%%%%%%%%%%%%%%%%%%%%%%%%%%%%%%%%%%%%%%%%%%%%%%%%%%%%%%
\section{Embedding Surface and Total Gravitational Energy}\label{sec6}
%%%%%%%%%%%%%%%%%%%%%%%%%%%%%%%%%%%%%%%%%%%%%%%%%%%%%%
In this section, we investigate some physical properties of the proposed wormhole structures, including the embedding surface and the total gravitational energy.

In three-dimensional space, the embedded surface $Z(r)$ of the axially symmetric wormhole is given by \cite{MT88}
\begin{equation}
ds^2 = \left[1+\left(\frac{d Z(r)}{dr}\right)^2\right]dr^2+r^2 d\phi^2.\label{es}
\end{equation}

Aslo, at the equatorial slice $\theta=\pi/2$ with constant time $t$, the metric (\ref{Metric}) reads as
\begin{equation}
ds^2 = \left(1-\frac{\xi(r)}{r}\right)^{-1}dr^2+r^2d\phi^2.\label{es1}
\end{equation}

Thus, combining results (\ref{es}) and (\ref{es1}) leads to the following differential equation for the embedding surface $Z(r)$   
\begin{equation}
\frac{dZ(r)}{dr} = \pm\left(\frac{r}{\xi(r)}-1\right)^{-\frac{1}{2}}.\label{dz1}
\end{equation}

The above result indicates that $dZ(r)/dr$ diverges at the throat, causing $Z(r)$ to become vertical at the throat. However, the embedding surface $Z(r)$ can be expressed as
\begin{equation}
Z(r) = \pm \int_{r_0^+}^{r} \frac{dr}{\sqrt{r/\xi(r)-1}},\label{z1}
\end{equation}
where the $\pm$ sign denotes the upper and lower regions of the wormhole geometry. Another key physical quantity of the wormhole is the proper radial distance, which we have already defined in Eq. (\ref{l}). 

The embedding surfaces and proper radial distances for the King and DZ DM models under model-I of $f(R, L_m)$ gravity are illustrated in Figs.~\ref{fig8} and \ref{fig9}, respectively. Fig. \ref{fig10} illustrates the embedding surface and proper radial distance for the model-II of $f(R, L_m)$ gravity.   In the embedding surface, the regions with positive curvature ($Z(r)>0$) and negative curvature ($Z(r)<0$) correspond to the wormhole's upper and lower universes, respectively, with the two joined at the wormhole throat. Moreover, the $f(R, L_m)$ gravity parameter $\alpha$ has a notable effect on the curvature of the wormholes; increasing $\alpha$ diminishes the spacetime curvature of the wormholes. The proper radial distance for the current wormhole solutions is finite and increases monotonically, also satisfying the condition $|l(r)| \geq r-r_0$. Furthermore, complete visualizations of the wormholes for each case are shown in Figs.~\ref{fig8}–\ref{fig10}. These analyses offer valuable insights into the physical viability of the proposed wormhole models within the $f(R, L_m)$ gravity framework.

 Now, we proceed to examine the characteristics of the total gravitational energy for the present wormhole models. The total gravitational energy, denoted by $E_g$, is defined as \cite{dl07, kk09}
\begin{eqnarray}
  E_g = Mc^2 -E_M.\label{eg}  
\end{eqnarray}
The term $Mc^2$ represents the total energy, defined as
\begin{eqnarray}
    Mc^2 = \frac{1}{2}\int_{r_0}^rT_t^tr^2dr+\frac{r_0}{2},\label{mc}
\end{eqnarray}

where the term $\frac{r_0}{2}$ denotes the effective mass \cite{kk09}.  The term $E_M$ denotes the total contribution from supplementary energy components, including rest, internal, kinetic, and related energies, and is expressed as
\begin{eqnarray}
    E_M = \frac{1}{2}\int_{r_0}^r \frac{r^2 T_t^t dr}{\sqrt{1-\xi(r)/r}}.\label{em} 
\end{eqnarray}

Thus, combining the results (\ref{mc}) and (\ref{em}), the total gravitational energy (\ref{eg}) can be expressed as
\begin{eqnarray}
  E_g = \frac{1}{2}\int_{r_0}^r\left[1-\frac{1}{\sqrt{1-\xi(r)/r}}\right]T_t^tr^2dr+\frac{r_0}{2}.\label{eg1}  
\end{eqnarray}

 The total gravitational energy is considered attractive when $E_g < 0$ and repulsive when $E_g > 0$ \cite{cw73}. For the present wormhole models, we demonstrate the graphical representation of the total gravitational energy in Fig.~\ref{fig11}, indicating $E_g > 0$. This suggests a repulsive gravitational effect, which supports the potential existence of viable traversable wormholes within the King and DZ DM halos under the $f(R, L_m)$ gravity framework.

%%%%%%%%%%%%%%%%%%%%%%%%%%%%%%%%%%%%%%%%%%%%%%%%%%%%%%
\section{Strong Deflection Angle}\label{sec7}
%%%%%%%%%%%%%%%%%%%%%%%%%%%%%%%%%%%%%%%%%%%%%%%%%%%%%%

In this section, we explore how the wormhole geometries considered in our study affect the path of light rays propagating in their vicinity. Specifically, we focus on calculating the strong gravitational deflection angle, which quantifies how much light is bent when it passes close to the wormhole throat. This analysis is crucial for understanding the gravitational lensing effects of these wormholes, which can provide observable signatures distinguishing them from black holes or other compact objects. Indeed, strong gravitational lensing provides a powerful tool for probing the geometry of spacetime and the properties of astrophysical objects \cite{ks98, cm01}. In this context, Bozza \cite{vb02} introduced a systematic analytical approach for examining light deflection in spherically symmetric spacetimes subjected to strong gravitational fields. The strong deflection angle $\alpha(r_c)$ corresponding to the wormhole metric (\ref{Metric}) can be written as \cite{vb01, vb02}
\begin{eqnarray}
\alpha(r_c)=-\pi+2\int_{r_{c}}^{\infty}\frac{e^{\Phi(r)}dr}{r^2\sqrt{\left(1-\frac{\xi(r)}{r}\right)\left(\frac{1}{\chi^{2}}-\frac{e^{2\Phi(r)}}{r^{2}}\right)}}, \label{DA}
\end{eqnarray}
where $r_c$ denotes the closest approach of the light ray to the wormhole throat, and $\chi$ represents the impact parameter. For null geodesics, these quantities are related as $\chi=r_c e^{-\Phi(r_c)}$.

The obtained deflection angles for the proposed wormhole structures are presented in Fig.~\ref{fig12} as a function of the closest approach distance $r_c$. The results indicate that the deflection angle diminishes as $r_c$ increases, i.e. weaker light bending at the greater distances from the wormhole throat. In contrast, near the throat, the deflection angle grows sharply and diverges, highlighting the intense curvature and strong gravitational influence in that region. Furthermore, the $f(R, L_m)$ gravity parameter $\alpha$ plays a significant role: larger $\alpha$ values reduce the deflection angle, indicating a decrease in gravitational lensing strength consistent with the reduced spacetime curvature.

%%%%%%%%%%%%%%%%%%%%%%%%%%%%%%%%%%%%%%%%%%%%%%%%%%%%%%%%%%%%%%%%%%%%%%%%%%%
\section{Discussions and Conclusion}\label{sec8}
%%%%%%%%%%%%%%%%%%%%%%%%%%%%%%%%%%%%%%%%%%%%%%%%%%%%%%%%%%%%%%%%%%%%%%%%%%%

Modified gravity theories offer an alternative explanation for the late-time cosmic acceleration. Among them, $f(R)$ gravity, where $f(R)$ is an arbitrary function of the Ricci scalar $R$, is particularly significant. More recently, the generalized $f(R, L_m)$ gravity, depending on both $R$ and the matter Lagrangian, has been introduced, with extensive astrophysical and cosmological implications. Recently, several investigations have demonstrated that $f(R, L_m)$ gravity offers a suitable theoretical framework for constructing and sustaining traversable wormhole geometries \cite{ns23, rs23, kd25, mm25, ae24}. It is noteworthy that, in standard Einstein gravity, sustaining wormhole geometries generally necessitates exotic matter \cite{ms88a}. In contrast, numerous studies have demonstrated that within modified gravity frameworks, wormhole solutions can often be constructed without the need for exotic matter \cite{mk15, ns25, gc12, mf20}. In this study, we have explored asymptotically flat non-exotic traversable wormhole structures within the King and DZ DM halos in the framework of $f(R, L_m)$ gravity under the consideration of two functional forms of this theory, Model-I: $f(R, L_{m})=(R/2)+L_{m}^{\alpha}$ and Model-II: $f(R, L_{m})=(R/2)+(1+\lambda R)L_{m}$. The key features of the proposed wormhole solutions can be summarized as follows:
\begin{itemize}
    \item \textbf{For Model-I wormholes:} In Model-I of $f(R, L_m)$ gravity, the wormhole solutions are sustained by two different DM density profiles, the King model and the Dekel–Zhao model. The King model generated shape function (\ref{B1}) increases monotonically with $r$ and decreases with higher $\alpha$, while satisfying $\xi(r)/r \leq 1$ for $r \geq r_0$ and fulfilling the flare-out condition for the parameters $\alpha = 2.2, 2.3, 2.4, 2.5, 2.6$ with $\beta = 0.65$, $\gamma = 1$, $\eta = -0.5$, $r_s = 1.01$, and $r_0 = 1.4$ (See Fig. \ref{fig1}). These behaviors confirm that the King DM based shape function encapsulates all the fundamental characteristics of a traversable wormhole. Furthermore, its asymptotic flatness indicates that the resulting wormhole geometries are asymptotically flat. To examine the physical characteristics of the matter supporting the wormholes, we analyze the relevant energy conditions. It is found that NEC$_r$, NEC$_t$, WEC$_r$, WEC$_t$, DEC$_r$, and SEC are satisfied throughout the wormhole spacetime, while DEC$_t$ holds near the throat (See Fig. \ref{fig2}). This indicates the absence of exotic matter, confirming that the King DM halo can sustain traversable wormholes in $f(R, L_m)$ gravity without exotic matter. The shape function (\ref{B2}) derived from the Dekel–Zhao model is also monotonically increasing with $r$, satisfies $\xi(r)/r \leq 1$ for $r \geq r_0$, meets the flare-out condition, and $\xi(r)/r \rightarrow 0$ as $r \rightarrow \infty$ for $\alpha = 2.2, 2.3, 2.4, 2.5, 2.6$ with $\rho_s = 0.06$, $\kappa = 2.12$, $r_s = 6$, and $r_0 = 1.4$, as shown in Fig. \ref{fig3}. Therefore, the obtained shape function, in this case, is also well-suited for constructing asymptotically flat traversable wormhole geometries. Moreover, these wormhole structures are supported by the non-exotic matter by satisfying all the relevant energy conditions, clear from Fig. \ref{fig4}.

    \item \textbf{For Model-II wormholes:} In Model-II of $f(R, L_m)$ gravity, we have introduced a new shape function (\ref{B3}) to construct the corresponding wormhole solutions. The newly proposed shape function, depicted in Fig.~\ref{fig5}, exhibits a monotonic increasing behavior with respect to the radial coordinate $r$, while $\xi(r) - r$ intersects the $r$-axis at $r = r_0 = 1.4$. It consistently satisfies the condition $\xi(r)/r \leq 1$ for $r \geq r_0$ and meets the flare-out requirement at the throat. These features confirm that the shape function effectively generates traversable wormhole geometries, fulfilling all necessary conditions. Additionally, its asymptotically flat nature ensures that the resulting spacetime remains asymptotically flat. These wormhole configurations within the King DM halo are supported by non-exotic matter by holding all the energy conditions, NEC, WEC, DEC, and SEC throughout the spacetime for the parameter set $\lambda = 0.11, 0.13, 0.15, 0.17, 0.19$ with $\beta = 0.65$, $\gamma = 1$, $\eta = -0.5$, $r_s = 1.01$, and $r_0 = 1.4$, as illustrated in Fig.~\ref{fig6}. Moreover, the wormhole configurations can also be sustained within the DZ DM halo without invoking exotic matter, as all energy conditions are satisfied for the parameter set $\lambda = 0.11, 0.13, 0.15, 0.17, 0.19$ with $\rho_s= 0.06$, $\kappa = 2.12$, $r_s$ = 6.5, and $r_0 = 1.4$, as shown in Fig.~\ref{fig7}. 
\end{itemize}

We have examined some physical characteristics of the proposed wormhole geometries to gain a deeper understanding of their structure. In particular, we have analyzed the embedding surface and proper radial distance, which provide a geometric visualization of the wormhole throat and its curvature properties in an embedded Euclidean space. The embedding surfaces and proper radial distances for the King and DZ DM models under Model-I of $f(R, L_m)$ gravity are shown in Figs.~\ref{fig8} and \ref{fig9}, while Fig.~\ref{fig10} presents the corresponding results for Model-II. In the embedding diagrams, the regions with positive curvature ($Z(r)>0$) and negative curvature ($Z(r)<0$) represent the upper and lower universes of the wormhole, respectively, joined smoothly at the throat. The $f(R, L_m)$ gravity parameter $\alpha$ significantly influences the wormhole geometry, with higher $\alpha$ values leading to reduced spacetime curvature. The proper radial distance for all considered models is finite, increases monotonically, and satisfies $|l(r)| \geq r-r_0$. Furthermore, we have analyzed the total gravitational energy to determine the nature of the gravitational interaction, whether attractive or repulsive, within the wormhole configurations. Interestingly, the results reveal that the total gravitational energy is repulsive in nature, as clear from Fig. \ref{fig11}, which supports the potential existence of viable traversable wormholes within the King and DZ DM halos under the $f(R, L_m)$ gravity framework.  Overall, these analyses provide deeper insight into the geometric and physical viability of the proposed wormhole configurations within the $f(R, L_m)$ gravity framework. In addition, we have investigated gravitational lensing, emphasizing the strong deflection effects generated by the wormhole geometries in each of the considered $f(R, L_m)$ gravity models, employing Bozza’s formalism \cite{vb02}. Our findings reveal that the deflection angles associated with the King and DZ DM profiles in Model-I $f(R, L_m)$ gravity and the introduced shape function in  Model-II $f(R, L_m)$ gravity exhibit a decreasing trend with increasing closest approach distance $r_c$ (See Fig. \ref{fig12}). The analysis implies weaker light bending farther from the wormhole throat. Near the throat, however, the deflection angle increases rapidly and diverges, indicating strong spacetime curvature and intense gravitational effects, where light can be trapped in unstable circular orbits. Moreover, the $f(R, L_m)$ gravity parameter $\alpha$ has a notable impact on the deflection angles; larger values of $\alpha$ lead to smaller deflection angles, reflecting a reduction in gravitational lensing strength due to the diminished curvature of spacetime.

In recent years, observations have confirmed black holes at galactic centers, but no direct evidence for wormholes exists. Consequently, the scientific community continue searching for observational signatures of wormholes. Following the pioneering work of Morris and Thorne, numerous theoretical models have been developed within GR and modified gravity theories, often at galactic scales. In this study, we have constructed new asymptotically flat traversable wormhole solutions within the framework of $f(R, L_m)$ gravity using the King and DZ DM models, without invoking exotic matter. Remarkably, these wormhole configurations satisfy all essential geometric and physical conditions, demonstrating their viability within the King and DZ DM halos under $f(R, L_m)$ gravity. Additionally, we have analyzed the corresponding deflection angles for each DM model, providing novel insights into gravitational lensing effects in the context of modified gravity. It is important to emphasize that these wormhole solutions are purely theoretical, and ongoing efforts in the scientific community aim to identify observational signatures, such as through scalar wave scattering, gravitational lensing \cite{pk14}, and gamma-ray burst light curve analyses \cite{dt98}. Furthermore, our study may inspire future research on constructing non-exotic traversable wormhole geometries within the King and DZ DM halos under other modified gravity theories, thereby extending their applicability across a wide range of alternative gravitational frameworks.

 \section*{Acknowledgments}
FR would like to thank the authorities of the Inter-University Centre for Astronomy and Astrophysics, Pune, India for providing research facilities.


\begin{thebibliography}{37}

\bibitem{POT1} L. Flamm, Physikalische Zeitschrift {\bf 17}, 448 (1916).

\bibitem{POT2} A. Einstein, and N. Rosen, Phys. Rev. {\bf 48}, 73 (1935). 

\bibitem{POT3} J. A. Wheeler, Ann. Phys. {\bf 2}, 604 (1957).

\bibitem{POT4} R. W. Fuller, and J. A. Wheeler, Phys. Rev. {\bf 128}, 919 (1962). 

\bibitem{POT5} M. S. Morris, and K. S. Thorne, Am. J. Phys. {\bf 56}, 395 (1988). 

\bibitem{POT6} M. Visser, {\tt Lorentzian Wormholes: From Einstein to Hawking},Springer, (1995).

\bibitem{POT7} J. A. Wheeler, Phys. Rev. {\bf 97}, 511 (1955).

\bibitem{POT8} M. Visser, S. Kar, and N. Dadhich, Phys. Rev. Lett. {\bf 90}, 201102 (2003).

\bibitem{POT9} F. S. N. Lobo and M. A. Oliveira, Phys. Rev. {\bf D 80}, 104012 (2009). 

\bibitem{POT10} T. Azizi, Int. J. Theor. Phys. {\bf 52}, 3486 (2013). 

\bibitem{POT11} S. H. Mazharimousavi, and M. Halilsoy, Mod. Phys. Lett. {\bf A 31}, 1650192 (2016). 

\bibitem{POT12} P. Moraes and P. Sahoo, Phys. Rev. {\bf D 96}, 044038 (2017). 

\bibitem{POT13} A. K. Mishra, U. K. Sharma, V. C. Dubey, and A. Pradhan, Astrophys. Space Sci. {\bf 365}, 34 (2020). 

\bibitem{POT14} A. Chanda, S. Dey, and B. C. Paul, Gen. Rel. Grav. {\bf 53}, 78 (2021). 

\bibitem{POT15} K. A. Bronnikov and S. W. Kim, Phys. Rev. {\bf D 67}, 064027 (2003).

\bibitem{POT16} M. L. Camera, Phys. Lett. {\bf B 573}, 27 (2003). 

\bibitem{POT17} F. Parsaei and N. Riazi, Phys. Rev. {\bf D 91}, 024015 (2015). 

\bibitem{POT18} F. Javed, G. Mustafa, A. Övgün, and M. F. Shamir, Eur. Phys. J. Plus {\bf 137}, 61 (2021). 

\bibitem{POT19} I. P. Lobo, M. G. Richarte, J. P. M. Graça, and H. Moradpour, Eur. Phys. J. Plus {\bf 135}, 550 (2020). 

\bibitem{POT20} T. Tangphati, C. Muniz, A. Pradhan, and A. Banerjee, Phys. Dark Univ. {\bf 42}, 101364 (2023). 

\bibitem{POT21} G. Mustafa, Z. Hassan, P. Moraes, and P. Sahoo, Phys. Lett. {\bf B 821}, 136612 (2021). 

\bibitem{POT22} Z. Hassan, G. Mustafa, J. R. L. Santos, and P. K. Sahoo, Europhys. Lett. {\bf 139}, 39001 (2022). 

\bibitem{POT23} A. Banerjee, A. Pradhan, T. Tangphati, and F. Rahaman, Eur. Phys. J. {\bf C 81}, 1031 (2021). 

\bibitem{POT24} Z. Hassan, S. Ghosh, P. K. Sahoo, and V. S. H. Rao, Gen. Rel. Grav. {\bf 55}, 90 (2023). 

\bibitem{POT25} R. Solanki, Z. Hassan, and P. K. Sahoo, Chin. J. Phys. {\bf 85}, 74 (2023). 

\bibitem{POT26} M. F. Shamir, G. Mustafa, S. Waseem, and M. Ahmad, Commun. Theor. Phys. {\bf 73}, 115401 (2021).

\bibitem{POT27} R. Shaikh, Phys. Rev. {\bf D 98}, 064033 (2018). 

\bibitem{POT28} S. Bahamonde, U. Camci, S. Capozziello, and M. Jamil, Phys. Rev. {\bf D 94}, 084042 (2016). 

\bibitem{POT29} K. Jusufi, Phys. Rev. {\bf D 98}, 044016 (2018). 

\bibitem{POT30} K. Jusufi, M. Jamil, and M. Rizwan, Gen. Rel. Grav. {\bf 51}, 102 (2019). 

\bibitem{FLMG1} T. Harko, and F. S. N. Lobo, Eur. Phys. J. {\bf C 70}, 373 (2010). 

\bibitem{FLMG2} T. Harko, Phys. Lett. {\bf B 669}, 376 (2008).  

\bibitem{FLMG3} T. Harko, Phys. Rev. {\bf D 81}, 084050 (2010).  

\bibitem{FLMG4} T. Harko, Phys. Rev. {\bf D 81}, 044021 (2010).  

\bibitem{FLMG5} S. Nesseris, Phys. Rev. {\bf D 79}, 044015 (2009).  

\bibitem{FLMG6} V. Faraoni, Phys. Rev. {\bf D 76}, 127501 (2007).  

\bibitem{FLMG7} V. Faraoni, Phys. Rev. {\bf D 80}, 124040 (2009).  

\bibitem{FLMG8} V. Faraoni, {\tt Cosmology in Scalar-Tensor Gravity}, Kluwer Academic, Dordrecht, (2004).

\bibitem{FLMG9} O. Bertolami, J. Paramos, and S. Turyshev, arXiv:gr-qc/0602016 (2006).  

\bibitem{FLMG10} J. Wang, and K. Liao, Class. Quantum Grav. {\bf 29}, 215016 (2012).  

\bibitem{FLMG11} B. S. Goncalves, P. H. R. S. Moraes, and B. Mishra, Fortschr. Phys., 71(8), 2200153 (2023).  

\bibitem{FLMG12} R. Solanki, B. Patel, L. V. Jaybhaye, and P. K. Sahoo, Commun. Theor. Phys. {\bf 75}, 075401 (2023).  

\bibitem{FLMG13} L. V. Jaybhaye, R. Solanki, S. Mandal, and P. K. Sahoo, Universe {\bf 9}, 163 (2023).  
\bibitem{FLMG14} M. Zeyauddin, A. Dixit, and A. Pradhan, Int. J. Geom. Meth. Mod. Phys. {\bf 21}(09), 2450167 (2023).  

\bibitem{FLMG15} N. Myrzakulov, M. Koussour, A. H. A. Alnadhief, and A. Abebe, Eur. Phys. J. Plus {\bf 138}, 852 (2023).  

\bibitem{FLMG16} D. C. Maurya, Grav. Cosm. {\bf 29}, 315 (2023).  

\bibitem{FLMG17} R. Solanki, et al., Commun. Theor. Phys. {\bf 75}, 075401 (2023).  

\bibitem{FLMG18} J. K. Singh, et al., New Astron. {\bf 104}, 102070 (2023).  

\bibitem{FLMG19} L. V. Jaybhaye, et al., Phys. Dark Univ. {\bf 40}, 101223 (2023).  

\bibitem{FLMG20} J. C. Fabris, et al., Eur. Phys. J. Plus {\bf 138}, 232 (2023).  

\bibitem{FLMG21} A. Pradhan, et al., Int. J. Geom. Meth. Mod. Phys. {\bf 20}, 1230105 (2023).  

\bibitem{FLMG22} D. C. Maurya, New Astron. {\bf 100}, 101974 (2023).  

\bibitem{FLMG23} G. A. Carvalho, et al., Eur. Phys. J. {\bf C 82}, 1096 (2022).  

\bibitem{FLMG24} R. V. Labato, G. A. Carvalho, and C. A. Bertulani, Eur. Phys. J. {\bf C 81}, 1013 (2021).

\bibitem{FLMG25} T. Harko, and S. Shahidi, Eur. Phys. J.  {\bf C 82}, 1003 (2022). 

\bibitem{FLMG26} T. Harko, and M. J. Lake, Eur. Phys. J. {\bf C 75}, 60 (2015).  


\bibitem{DM2} K. K. Nandi, Y.-Z. Zhang, and K. B. Vijaya Kumar, Phys. Rev. {\bf D 70}, 127503 (2004).

\bibitem{DM3} M. S. Churilova, R. A. Konoplya, Z. Stuchlik, and A. Zhidenko, JCAP {\bf 10}, 010 (2021).

\bibitem{DM4} R. A. Konoplya and A. Zhidenko, Phys. Rev. Lett. {\bf 128}, 091104 (2022).

\bibitem{DM5} A. Ashtekar, J. Phys. Conf. Ser. {\bf 189}, 012003 (2009).

\bibitem{DM6} R. Sengupta, S. Ghosh, and M. Kalam, Eur. Phys. J. {\bf C 83}, 830 (2023).

\bibitem{DM7} C. R. Muniz, T. Tangphati, R. M. P. Neves, and M. B. Cruz, Phys. Dark Univ. {\bf 46}, 101673 (2024).

\bibitem{DM8} N. Aghanim, et al. (Planck), Astron. Astrophys. {\bf 641}, A6 (2020).

\bibitem{DM9} F. Zwicky, Helv. Phys. Acta {\bf 6}, 110 (1933).

\bibitem{DM10} V. C. Rubin and W. K. Ford, Jr., Astrophys. J. {\bf 159}, 379 (1970).

\bibitem{DM11} M. Persic, P. Salucci, and F. Stel, Mon. Not. Roy. Astron. Soc. {\bf 281}, 27 (1996).

\bibitem{DM12} G. Bertone and D. Hooper, Rev. Mod. Phys. {\bf 90}, 045002 (2018).

\bibitem{DM13} L. Randall, Nature {\bf 557}, 2 (2018).

\bibitem{B1} A. Ashraf, et al.,  Phys. Dark Universe, {\bf 47}, 101787 (2025).

\bibitem{B2} G. Mustafa, et al., Phys. Dark Universe, {\bf 47}, 101765 (2025).

\bibitem{B3} G. Mustafa, et al.,  Phys. Dark Universe, {\bf 47}, 101753 (2025).

\bibitem{B4} A. Ashraf, et al., Physics of the Dark Universe, {\bf 47}, 101725 (2025).

\bibitem{B5} A. Ashraf, et al., Physics of the Dark Universe {\bf 48}, 101836 (2025).

\bibitem{B6} A. Ashraf, et al., Phys. Dark Universe {\bf 47}, 101823 (2025).

\bibitem{B7} A. Ditta, et al.,  Phys. Dark Universe {\bf 47}, 101818 (2025). 

\bibitem{B8} A. Ditta, et al.,  Phys. Dark Universe, {\bf 46}, 101573 (2024).

\bibitem{B9} A. Bouzenada, et al., Nucl. Phys. {\bf B 1017}, 116928 (2025).

\bibitem{B10} A. Saleem, et al., Nucl. Phys. {\bf B 1017}, 116926 (2025).
 
\bibitem{DM14} C. A. Argüelles, et al., Rev. Mod. Phys. {\bf 93}, 035007 (2021).

\bibitem{DM15} D. J. E. Marsh, D. Ellis, and V. M. Mehta, {\tt Dark Matter: Evidence, Theory, and Constraints, Princeton Series in Astrophysics}, Princeton University Press, (2024).

\bibitem{DM16} Y.-D. Tsai, J. Eby, and M. S. Safronova, Nature Astron. {\bf 7}, 113 (2023).

\bibitem{DM17} A. D. S. Souza, C. R. Muniz, R. M. P. Neves, and M. B. Cruz, Ann. Phys. {\bf 472}, 169859 (2025).

\bibitem{DM18} Z. Xu, M. Tang, G. Cao, and S.-N. Zhang, Eur. Phys. J. {\bf C 80}, 70 (2020).

\bibitem{DM19} C. R. Muniz and R. V. Maluf, Annals Phys. {\bf 446}, 169129 (2022).

\bibitem{DM20} G. Mustafa, S. K. Maurya, and S. Ray, Fortsch. Phys. {\bf 71}, 2200129 (2023).

\bibitem{DM21} R. Radhakrishnan, et al., Symmetry {\bf 16}, 1007 (2024).

\bibitem{DM22} A. Errehymy, et al., Eur. Phys. J. {\bf C 84}, 573 (2024).

\bibitem{DM23} S. K. Maurya, J. Kumar, S. Kiroriwal, and A. Errehymy, Phys. Dark Univ. {\bf 46}, 101564 (2024).

\bibitem{DM24} Z. Hassan, and P. K. Sahoo, Annalen Phys. {\bf 536}, 2400114 (2024).

\bibitem{DM25} Z. Xu, M. Tang, G. Cao, and S.-N. Zhang, Eur. Phys. J. {\bf C 80}, 70 (2020).

\bibitem{DM26} J. F. Navarro, C. S. Frenk, and S. D. M. White, Astrophys. J. {\bf 462}, 563 (1996).

\bibitem{DM27} K. G. Begeman, A. H. Broeils, and R. H. Sanders, Mon. Not. Roy. Astron. Soc. {\bf 249}, 523 (1991).

\bibitem{King} P. H. Chavanis, M. Lemou, and F. Méhats, Phys. Rev. {\bf D 91}(6), 063531 (2015).

\bibitem{DZM1} H. Zhao, {\it Mon. Not. R. Astron. Soc.} \textbf{278}, 488 (1996).

\bibitem{DZM2} H. Zhao, {\it Mon. Not. R. Astron. Soc.} \textbf{287}, 525 (1997).

\bibitem{DZM3} A. Dekel, G. Ishai, A. A. Dutton, and A. V. Macciò, {\it Mon. Not. R. Astron. Soc.} \textbf{468}, 1005 (2017).

\bibitem{DZM4} J. Freundlich, et al., {\it Mon. Not. R. Astron. Soc.} \textbf{499}, 2912 (2020).

\bibitem{DZM5} P. Salucci, {\it Universe} \textbf{11}, 67 (2025).

\bibitem{DZM6} D. Batic, J. M. Faraji, and M. Nowakowski, {\it Eur. Phys. J. C} \textbf{ 82}, 759 (2022).

\bibitem{DZM7} A. A. Badawi, S. Shaymatov, and Y. Sekhmani, {\it JCAP} \textbf{02}, 014 (2025).

\bibitem{DZM8} A. Einstein, {\it Ann. Math.} \textbf{40}, 922 (1939).

\bibitem{DZM9} V. Cardoso, et al., {\it Phys. Rev. Lett.} \textbf{129}, 241103 (2022).

\bibitem{DZM10} L. Hernquist, {\it Astrophys. J.} \textbf{356}, 359 (1990). 

\bibitem{DZM11} M. Khatri, and P. K. Sahoo, {\it Phys. Dark Universe} {\bf 49}, 102042 (2025).

\bibitem{DZM12} A. Errehymy, O. Donmez, A. Syzdykova, K. Myrzakulov, S. Muminov, A. Dauletov, and J. Rayimbaev, {\it Ann. Phys.} {\bf 480}, 170105 (2025).)


\bibitem{th10} T. Harko and F. S. N. Lobo, {\it Eur. Phys. J. C} {\bf 70}, 373-379 (2010).


 
\bibitem{MT88} M. S. Morris, K.S. Thorne, {\it Am. J. Phys.} {\bf 56}, 395 (1988).
\bibitem{ms88a} M. S. Morris, K. S. Thorne and U. Yurtsever, {\it Phys. Rev. Lett.} {\bf 61}, 1446 (1988).
\bibitem{ar55} A. Raychaudhuri, {\it Phys. Rev.} {\bf 98}, 1123 (1955).
\bibitem{lv22} L. V. Jaybhaye et al., {\it Phys. Lett. B} {\bf 831}, 137148 (2022).
\bibitem{ir72} I. R. King, {\it ApJL} {\bf 174}, L123 (1972).
\bibitem{an18} A. N. Baushev, {\it New Astronomy} {\bf 60} 69-73 (2018).
\bibitem{DZ1} H. Zhao, {\it Mon. Not. R. Astron. Soc.} \textbf{278}, 488 (1996).
\bibitem{DZ2} H. Zhao, {\it Mon. Not. R. Astron. Soc.} \textbf{287}, 525 (1997).
\bibitem{DZ4} J. Freundlich, et al., {\it Mon. Not. R. Astron. Soc.} \textbf{499}, 2912 (2020).
\bibitem{nm10} N. M. Garcia, F. S. N. Lobo, {\it Phys. Rev. D} {\bf 82}, 104018 (2010).
\bibitem{nm11} N. M. Garcia, F.S.N. Lobo, {\it Class. Quantum Gravity} {\bf 28}, 085018
(2011).
\bibitem{rv21} R. V. Labato, G. A. Carvalho, and C. A. Bertulani, {\it Eur. Phys. J. C} {\bf 81}, 1013 (2021).
\bibitem{dl07} D. Lynden-Bell, J. Katz, and J. Bicak,  {\it Phys. Rev. D} {\bf 75}, 024040 (2007).
\bibitem{kk09} K. K. Nandi, Y. Z. Zhang, R. G. Cai, and A. Panchenko, {\it Phys. Rev. D} {\bf  79}, 024011 (2009).
\bibitem{cw73} C. W. Misner, K. S. Thorne, and J. A. Wheeler, {\tt Gravitation}, San Francisco, (1973).
\bibitem{ks98} K. S. Virbhadra, D. Narasimha, and S. M. Chitre, {\it Astron. Astro. Phys.} {\bf 337}, 1 (1998).
\bibitem{cm01} C. M. Claudel, K. S. Virbhadra, and G. F. R. Ellis, {\it J. Math. Phys.} {\bf 42}, 818 (2001).
\bibitem {vb02} V. Bozza, {\it Phys. Rev. D} {\bf 66}, 103001 (2002).
\bibitem {vb01} V. Bozza et al., {\it Gen. Relativ. Gravit.} {\bf 33}, 1535 (2001).


\bibitem {ns23} N. S. Kavya, et al., {\it Chinese Journal of Physics} {\bf 84} 1-11 (2023).
\bibitem {rs23} R. Solanki, Z. Hassan, and P. K. Sahoo, {\it Chinese Journal of Physics} {\bf 85}, 74-88 (2023).
\bibitem {kd25} K. De, S. Mitra, and U. C. De,  {\it International Journal of Geometric Methods in Modern Physics} {\bf 22} 2450265 (2025).
\bibitem {mm25} M. M. Rizwan, Z. Hassan, and P. K. Sahoo, {\it Physics Letters B} {\bf 860}, 139152 (2025).
\bibitem {ae24} A. Errehymy, {\it Chinese Journal of Physics} {\bf 89}, 56-68 (2024).
\bibitem{mk15} M. K. Zangeneh, F. S. N. Lobo and M. H. Dehghani, {\it Phys. Rev. D} {\bf 92}, 124049 (2015).
\bibitem{ns25} N. Sarkar, S. Sarkar, M. Sarkar and F. Rahaman, {\it Physics of the Dark Universe} {\bf 47}, 101828 (2025).
\bibitem{gc12} C. G. Boehmer, T. Harko and F. S. N. Lobo, {\it Phys. Rev. D} {\bf 85}, 044033 (2012).
\bibitem{mf20} M. F. Shamir, and I. Fayyaz, {\it Eur. Phys. J. C} {\bf 80}, 1-9 (2020)
\bibitem {pk14} P. K. F Kuhfittig, {\it Eur. Phys. J. C}  {\bf 74}, 2818 (2014).
\bibitem {dt98} D. Torres,  G. Romero, L. Anchordoqui,  {\it Phys. Rev. D} {\bf 58}, 123001 (1998).
\end{thebibliography}
\end{document}